\documentclass[showpacs,preprintnumbers,amsmath,amssymb,floatfix,nofootinbib]{revtex4}

\usepackage{amsmath,amsfonts,bbm,euscript,hyperref}
\usepackage{graphicx}
\usepackage{dcolumn}
\usepackage{bm}
\usepackage{epsfig}
\epsfclipon

\newcommand{\lsim}{\mbox{\raisebox{-.6ex}{~$\stackrel{<}{\sim}$~}}}
\newcommand{\gsim}{\mbox{\raisebox{-.6ex}{~$\stackrel{>}{\sim}$~}}}

\def\sH{\mathcal{H}}
\def\Lap{\triangle}

\begin{document}

\preprint{UMN-TH-3031/12}

\title{Observable non-gaussianity from gauge field production in slow roll inflation, and a challenging connection with magnetogenesis}

\author{Neil Barnaby, Ryo Namba, Marco Peloso}

\affiliation{
 School of Physics and Astronomy,
University of Minnesota, Minneapolis, 55455 (USA)\\
}

\date{\today}

\begin{abstract} 
In any realistic particle physics model of inflation, the inflaton can be expected to couple to other fields.  We consider a model with a dilaton-like coupling between a U(1) gauge field and a scalar inflaton.  We show that this coupling can result in observable non-gaussianity, even in the conventional regime where inflation is supported by a single scalar slowly rolling on a smooth  potential:  the time dependent inflaton condensate leads to amplification of the large-scale gauge field fluctuations, which can feed-back into the  scalar/tensor cosmological perturbations. In the squeezed limit, the resulting bispectrum is close to the local one, but it shows a sizable and characteristic quadrupolar dependence on the angle between the shorter and the larger modes in the correlation. Observable non-gaussianity is obtained in a regime where perturbation theory is under control. If the gauge field is identified with the electromagnetic field, the model that we study is a realization of the magnetogenesis idea originally proposed by Ratra, and widely studied. This identification (which is not necessary for the non-gaussianity production) is however problematic in light of a strong coupling problem already noted in the literature.
\end{abstract}

\maketitle

\begin{widetext}
\tableofcontents\vspace{5mm}
\end{widetext}

\section{Introduction} 
\label{sec:introduction}

Over the next few years Cosmic Microwave Background (CMB) and Large Scale Structure (LSS) probes will measure the primordial cosmological fluctuations with substantial improvements in accuracy and over a widening range of scales.  With this wealth of data there is an exciting prospect to strongly constrain, or perhaps measure, nongaussian statistics of the primordial density perturbations.  Nongaussian effects encode a wealth of information about the physics of the very early universe and might provide a powerful tool to discriminate between different models.  (See \cite{Barnaby:2010sq} for a recent review.)

It is often claimed that primordial nongaussianity will be undetectable in the simplest models, where the inflation is driven by a single field that is slowly rolling along a smooth, flat potential.  The physical reason is that nongaussianity is a measure of the strength of interactions, while the requirement of a flat potential usually constrains inflaton self-interactions  to be weak \cite{Acquaviva:2002ud,Maldacena:2002vr,Seery:2005wm,Seery:2008qj}.  A number of models have been constructed which \emph{do} produce a detectable nongaussian signature, for example using sound speed effects \cite{small_sound}, higher derivatives \cite{NL}, non-vacuum initial conditions \cite{small_sound,nonBD1,nonBD2,nonBD}, sharp potential features \cite{chen1,chen2}, post-inflationary effects \cite{preheatNG,preheatNG2}, etc.

However, in general there are not only inflaton self-couplings, or gravitationally suppressed couplings.  For instance, in any realistic particle physics framework, the inflaton field $\varphi$ can be expected to couple to ``matter'' fields.\footnote{Here we use the word ``matter'' to describe any fields that do not play a significant role in driving the inflationary expansion.}  Such couplings are certainly necessary for   successful reheating after inflation, and they are much less constrained by the requirement of slow roll. Their consistent inclusion can radically impact the phenomenology of the model; see the works \cite{Berera:1995ie,Anber:2009ua,Barnaby:2010vf,Barnaby:2011vw,Barnaby:2011qe,Barnaby:2009mc,Barnaby:2009dd,Barnaby:2010ke,Cook:2011hg,Sorbo:2011rz} for explicit examples and scenarios.  

We focus our attention to inflaton interaction with gauge fields. For a singlet inflaton, there are two very natural classes of gauge field interactions to consider, depending on the parity.  For a pseudoscalar inflaton one expects axial interactions of the type
\begin{equation}
\label{Lint_pseudo}
  \mathcal{L}^{\mathrm{pseudo}}_{\mathrm{int}} = -\frac{\varphi}{f} F^{\mu\nu} \tilde{F}_{\mu\nu} \, ,
\end{equation}
where $F_{\mu\nu} = \partial_\mu A_\nu - \partial_\nu A_\mu$ is the field strength associated to some $U(1)$ gauge field $A_\mu$ and $F^{\mu\nu} = \epsilon^{\mu\nu\alpha\beta} F_{\alpha\beta} / 2$ is its dual.  For a scalar inflaton, on the other hand, one may expect couplings of the form
\begin{equation}
\label{Lint_scalar}
  \mathcal{L}^{\mathrm{scalar}}_{\mathrm{int}} = - \frac{I^2(\varphi)}{4} F^{\mu\nu} F_{\mu\nu} \, ,
\end{equation}
where $I(\varphi)$ plays the role of a field dependent gauge coupling.  The interaction (\ref{Lint_scalar}) is typical of moduli or dilaton-like fields in string theory and supergravity frameworks. As noted in
\cite{Ferrara:2011dz}, coupling the inflaton to a gauge field is actually the only way to reheat for some of these models. 

Pseudoscalar couplings of the type (\ref{Lint_pseudo}) have been studied extensively in \cite{Barnaby:2010vf,Barnaby:2011vw,Barnaby:2011qe}.  For natural values of $f$, such couplings lead to a copious production of gauge field fluctuations that feed-back on the scalar and tensor cosmological perturbations.  Observable nongaussianity of the equilateral type is naturally generated, without any exotic model-building ingredients \cite{Barnaby:2010vf,Barnaby:2011vw}.  Moreover, this is correlated also with a gravitational wave signal that can be detectable with interferometers \cite{Cook:2011hg,Barnaby:2011qe}.  

In this paper, we show that a similarly rich phenomenology is possible also in simple models with a single \emph{scalar} inflaton in slow roll, via the coupling (\ref{Lint_scalar}).  The underlying mechanism is quite novel.  The time-dependence of the inflaton condensate $\varphi_0(t) = \langle \varphi(t,{\vec x})\rangle$ breaks the conformal invariance of the gauge field sector and leads to amplification of the quantum fluctuations of $A_\mu$, similarly to the well-known mechanism that produces scale invariant curvature fluctuations during inflation.  For simplicity, we focus our attention on the case where large scale fluctuations of the gauge field are produced during inflation with a scale-invariant ``magnetic'' component.  We notice that the \emph{same} coupling that leads to production of gauge field fluctuations also implies that these produced fluctuations must, in turn, couple to the cosmological perturbations of the inflaton, $\delta\varphi(t,{\vec x}) = \varphi(t,{\vec x}) - \varphi_0(t)$.  We find that the feed-back of produced gauge fluctuations on $\delta\varphi$ contributes a new component to the observable curvature fluctuations that is highly nongaussian and is uncorrelated with the usual spectrum from quantum vacuum fluctuations.  For reasonable parameters, we obtain nearly local-type nongaussianity
with shape
\begin{equation}
\langle \zeta_{{\vec k}_1} \,  \zeta_{{\vec k}_2} \,  \zeta_{{\vec k}_3} \rangle \propto
\frac{1+\cos^2 \left( \vec{k}_1 ,\, \vec{k}_2 \right)}{k_1^3 \, k_2^3} + {\rm permutations}
\label{shape-intro}
\end{equation}
and at the level $f_{NL} = \mathcal{O}(10-100)$.  Such values are close to current observational limits and will be probed in the near future.  The shape (\ref{shape-intro}) has a strong overlapping with the local template in the squeezed limit, where both shapes are enhanced. However, in this limit, (\ref{shape-intro}) has a quadrupolar dependence on the angle between the shorter side and either of the longer sides. Contrary to what typically happens for non-gaussianity sourced by scalar fields \cite{Lewis:2011au}, this angular dependence does not vanish in the squeezed limit, and it contributes to about $1/5$ of the amplitude of (\ref{shape-intro}). This therefore appears as a signature of non-gaussianity from higher spin fields, and it may be an important distinguishing feature when the model is confronted with observations.~\footnote{We thank Antony Lewis for stressing the importance of this in a private communication.}

Our scenario is a realization of the ``feeder'' mechanism \cite{Barnaby:2011pe}; consequently, the Probability Distribution Function (PDF) exhibits a non-hierarchical structure.  This unique feature could help to observationally distinguish the model from other constructions that give nearly local nongaussianity (for example the curvaton \cite{Lyth:2001nq,Bartolo:2003jx}).

Models of vector fields producing non-gaussianity were also proposed in \cite{vector-curvaton}. In 
these models the vector is amplified either from the time dependence of  its kinetic term, as we consider here, or from a nonminimal coupling (as an effective mass term) to the scalar curvature $R$. This second case, however, introduces a longitudinal vector component, which turns out to be a ghost  \cite{Himmetoglu:2008zp,Himmetoglu:2009qi}. (In some of these models, the energy in the vector field sources anisotropic inflation; we do not study this possibility here, but we refer the reader to Ref. \cite{Soda:2012zm} for a review.)  In these models, the vector acts as a curvaton, and it contributes to the spectrum and the bispectrum through its energy density. We compute instead the non-gaussianity resulting from the  \emph{same} coupling that leads to production of gauge field fluctuations, and which remains imprinted in the inflaton perturbations even when the energy density in the gauge field is very negligible at reheating. The model of ref. \cite{Yokoyama:2008xw} studied non-gaussianity from a gauge field amplified analogously to what we do here, but imprinted 
 through the waterfall field of hybrid inflaton \cite{Linde:1993cn}.

It is interesting to contrast the mechanism considered in this paper to the closely related physics of inverse decay that was studied in \cite{Barnaby:2010vf,Barnaby:2011vw}.  In the case at hand, the feed-back of the produced gauge field fluctuations on the inflaton fluctuations leads to a super-horizon growth of the curvature perturbation.  Such growth is consistent, since the produced gauge field fluctuations provide a source of large scale iso-curvature perturbations.  Consequently, we find a bispectrum which is very close to the local shape.  On the other hand, for axial couplings one finds \cite{Barnaby:2010vf} that the relevant production of inflaton perturbations arises near horizon crossing, and then the bispectrum is instead nearly equilateral.

Models of the type which we study have received considerable attention in the literature, in connection with primordial magnetogenesis \cite{Ratra:1991bn,Gasperini:1995dh,Widrow:2002ud,Bamba:2003av,Martin:2007ue,Kanno:2009ei,Subramanian:2009fu,Seery:2008ms,Demozzi:2009fu,Caldwell:2011ra}.  At face value, our choice of $I^2(\varphi)$ can produce large scale magnetic fields with a sufficient amplitude to account for observations at galaxy and cluster scales; see \cite{Grasso:2000wj,Giovannini:2003yn,Widrow:2011hs,Ryu:2011hu,Kandus:2010nw} for reviews. This would open the interesting possibility of  correlating the magnetic field with the primordial perturbations \cite{Caldwell:2011ra}, although the correlation would only involve the component of metric perturbations that are sourced by the vector field, and that is typically subdominant with respect to the vacuum part. Moreover, the magnetic field would induce non-gaussianiy from its direct coupling to the CMB photons \cite{B-zzz}. This effect can be observed for a magnetic field at the $\sim {\rm nG} - 10 \, {\rm nG}$ level, while the non-gaussianity we have obtained arises from the direct coupling to the inflaton that generated the gauge field, and can be observable even if the current ``magnetic'' field is significantly smaller. A list of works that study the general signatures of a magnetic field on the CMB can be found in the review \cite{Widrow:2011hs}. In particular, ``magnetic'' fields continue to source scalar and tensor perturbations until neutrino decoupling. Refs. \cite{Shaw-Trivedi} shows that this effect constrains the ``magnetic'' field to be $\lsim {\rm few \, nG}$ at present. We show in Subsection \ref{subsec:magnetic} that the ``magnetic field'' generated in the cases of our interest is of ${\rm O } \left( 10^{-10} \right) \, {\rm nG}$ or less.

One major problem with identifying the gauge field with the electromagnetic one is due to the fact that a scale invariant magnetic field can only be obtained if  the effective gauge coupling, $g(t) = I\left[\varphi_0(t)\right]^{-1}$, decreases by \emph{many} orders of magnitude during inflation (specifically, by a factor of ${\rm e}^{2 N_{\rm tot}}$, where $N_{\rm tot}$ is the number of e-folds of inflation).

If one starts with from $g_{\mathrm{in}} = \mathcal{O}(1)$, then the gauge coupling at the end of inflation will be \emph{much} too small to identify $A_\mu$ with the Standard Model (SM) photon. Normalizing instead the coupling constant to be the electromagnetic one at the end of inflation would instead  entail an unacceptable breakdown of perturbation theory. This problem was  stressed in \cite{Demozzi:2009fu}, and it is a serious obstacle in identifying the gauge field of the mechanism with the photon. In the following, we discuss some unsuccessful attempts of solving this problem. We cannot of course rule out that a solution of the problem can be found, but we believe that it would require a substantial modification of the model.
 
As a consistency check on our calculation, we have verified that perturbation theory is well under control using a variety of different diagnostics, including backreaction effects, the amplitude of curvature perturbations, the energy density in fluctuations, and the Weyl tensor.  We show that the ``new'' metric 
perturbations which are sourced by the gauge field fluctuations are generically sub-dominant, as compared to the standard vacuum contribution.  

The organization of this paper is as follows.  In Section \ref{sec:EB} we introduce our model and compute the production of gauge field fluctuations. We then discuss the challenging connection to  magnetogenesis.  In Section \ref{sec:pheno} we give a succinct review of the key phenomenological predictions for the spectrum and bispectrum of curvature fluctuations, and also the spectrum of tensor fluctuations.    In Section \ref{sec:scalar} we present a detailed computation of the 2-point and 3-point correlation functions of the scalar perturbations.  In Section \ref{sec:tensor} we present the computation of the 2-point correlation function of gravitational wave fluctuations.  In Section \ref{sec:weyl} we discuss the validity of our perturbative analysis.  Finally, in Section \ref{sec:conclusions}, we conclude.  Appendix A gives some technical details on second order cosmological perturbation theory in the spatially flat gauge.  In this paper we compute the feed-back of the produced gauge field fluctuations on the inflaton using the Green function method that was developed in \cite{Barnaby:2009mc,Barnaby:2010ke} and employed also in \cite{Barnaby:2010vf,Barnaby:2011vw}.  This formalism is equivalent to the ``in-in'' approach at leading order.  In Appendix B we demonstrate this equivalence explicitly for the case at hand.  (See also \cite{Barnaby:2011qe}.)

\section{Gauge Field Production}
\label{sec:EB}

\subsection{The Model}

We consider a simple model with a dilaton-like coupling of the inflaton to a $U(1)$  gauge field $A_\mu$
\begin{equation}
S = \int d^4 x \sqrt{-g} \left[ \frac{M_p^2}{2} R - \frac{1}{2} \left( \partial \varphi \right)^2 - V \left( \varphi \right) - \frac{I^2 \left( \varphi \right)}{4} F^2 \right]
\label{model1}
\end{equation}
In this action, $F_{\mu \nu}$ is the field strength of $A_\mu$, and  $I$ and $V$ are functions of the inflaton $\varphi$.  We assume that the potential $V(\varphi)$ is sufficiently flat to support a long phase of quasi de Sitter expansion.  As usual, we require that $\epsilon,|\eta|\ll 1$ where the slow roll parameters are
\begin{equation}
 \epsilon \equiv \frac{M_p^2}{2}\left(\frac{V_{,\varphi}}{V}\right)^2 \, \hspace{5mm} \eta \equiv M_p^2 \frac{V_{,\varphi\varphi}}{V}
\end{equation}

To simplify the calculation, we choose $I(\varphi)$ and $V(\varphi)$ to be related to each other in such a way as to obtain the solution $I \propto a^n$, where $a$ is the scale factor of the Universe, during inflation.  The required relation can be obtained  \cite{Watanabe:2009ct}  by taking the ratio of the slow roll relations
\begin{equation}
\label{SReqns}
H^2 \simeq \frac{1}{3 M_p^2} V \;\;,\;\; 3 \, H \dot{\varphi} \simeq - V_{,\varphi}
\end{equation}
(assuming that the interaction with the gauge field provides negligible corrections to the dynamics, see below). Here and in the following, dot denotes derivative with respect to physical time,  ``${,\varphi}$'' denotes derivative with respect to $\varphi$, while $H$ is the Hubble rate, $H = \frac{\dot{a}}{a}$. In this way, one forms a differential equation for $d a / d \varphi$, that is integrated into $a \propto {\rm exp} \left[ - \int \frac{V d \varphi}{V_{,\varphi} M_p^2} \right]$. The functional form of $I \left( \varphi \right)$ can be then set to the  $n-th$ power of the right hand side expression. For definiteness, we consider a monomial inflaton potential
\begin{equation}
V = \mu^{4-r} \varphi^r \;\;,\;\; I = I_{\mathrm{end}} \exp\left[-\frac{n \varphi^2}{2 r M_p^2}\right]
\label{model2}
\end{equation}
although other choices are clearly possible.  Here $I_{\mathrm{end}}$ is the value of the coupling function at the end of inflation, when the inflaton is in the vacuum $\varphi=0$.  We therefore assume that, after inflation, the function $I$ sets to a constant. We could then normalize $I_{\rm end} = 1$.

We found it algebraically convenient to define ``electric'' and ``magnetic'' components of the gauge field as
\begin{equation}
E_i \equiv - \frac{\langle I \rangle}{a^2} A_i' \;\;,\;\; B_i \equiv \frac{\langle I \rangle}{a^2}  \epsilon_{ijk} \partial_j A_k
\label{EB-def}
\end{equation}
where here and in the remainder of this work the Coulomb gauge $A_0 = 0$ is  assumed.  We do not necessarily  assume $A_\mu$ is the Standard Model photon (more on this later), however, we will sometimes use the language ``electric field'' and ``magnetic field'', by analogy with standard electromagnetism.  

As the gauge field has no classical expectation value, its perturbations do not couple to that of the inflaton or of the geometry at linearized order. We can therefore solve for these perturbations by assuming a FRW background, and by treating $I$ as a classical function. Therefore, we can simply set $I\propto a^n$ for the remainder of this Section. The time component of the vector equation of motion is solved by $\partial_i A_i = 0$, and we can decompose the vector potential as
\begin{equation}
\vec{A} = \sum_{\lambda=\pm} \int \frac{d^3 k}{\left( 2 \pi \right)^{3/2}} 
\vec{\epsilon}_\lambda \left( \vec{k} \right)
 {\rm e}^{i \vec k \cdot \vec x} \, 
\left[  \, a_\lambda \left( \vec{k} \right) \, A_\lambda \left( k \right) + 
 a_\lambda^\dagger \left( - \vec{k} \right) \, A_\lambda^* \left( k \right)       \right] \;\;\;,\;\;\;
\label{A-deco}
\end{equation}
where the circular polarization operators satisfy $\vec{k}\cdot \vec{\epsilon}_{\pm} \left( \vec{k} \right) = 0$, $\vec{k} \times \vec{\epsilon}_{\pm} \left( \vec{k} \right) = \mp i k \vec{\epsilon}_{\pm} \left( \vec{k} \right)$, $\vec{\epsilon}_\pm \left( \vec{-k} \right) = \vec{\epsilon}_\pm \left( \vec{k} \right)^*$, and are  normalized according to $\vec{\epsilon}_\lambda \left( \vec{k} \right)^* \cdot \vec{\epsilon}_{\lambda'} \left( \vec{k} \right) = \delta_{\lambda \lambda'}$. The annihilation / creation operators satisfy
$\left[ a_\lambda \left( \vec{k} \right) ,\, a_{\lambda'}^\dagger \left(  \vec{k'} \right) \right] = \delta_{\lambda \lambda'} \, \delta^{(3)} \left( \vec{k} - \vec{k}' \right)$.

The vector mode functions then satisfy
\begin{equation}
\label{Veqn}
V_\lambda'' + \left( k^2 - \frac{I''}{I} \right) V_\lambda = 0 \;\;\;,\;\;\; V_\lambda \equiv I \, A_\lambda
\end{equation}
where prime denotes derivative with respect to conformal time $\tau$. For a constant $I $, one recovers the typical Minkowski ${\rm e}^{-i k \tau}/\sqrt{2 k}$ solution for the mode functions, due to the fact that the gauge field is conformally coupled to the FRW metric. For our purposes, it is sufficient to obtain the leading expression of the mode functions in a slow roll expansion. Namely, for the de Sitter  limit $a= - \frac{1}{H \tau}$,  the mode solution that, up to an arbitrary constant phase, is  normalized to ${\rm e}^{-i k \tau}/\sqrt{2 k}$ (the so called adiabatic vacuum) in the asymptotic early time / high momentum regime is
\begin{equation}
V_\lambda = i \frac{\sqrt{\pi}}{2} \, \sqrt{-\tau} \, H_{n+1/2}^{(1)} \left( - k \tau \right)
\label{full-sol}
\end{equation}
This solution has been discussed at length in \cite{Demozzi:2009fu} for all values of $n$, and it is unnecessary to review all their study here. We only discuss the $n=2$ case, which results in a scale invariant ``magnetic'' field. Interesting non-gaussian properties of the primordial perturbations may be possible also for other values of $n$. The arbitrary phase in (\ref{full-sol}) has been chosen so that the function $V_\lambda$ is real and positive in the super horizon limit.

In this case, the ``electric'' and ``magnetic'' gauge field operators during inflation reduce to
\begin{eqnarray}
\begin{array}{l}
\vec{E}  = \int \frac{d^3 k}{\left( 2 \pi \right)^{3/2}}  \, {\rm e}^{i \vec{k} \cdot \vec{x}} \, \vec{E}_{\vec k} \;\;,\;\;
\vec{E}_{\vec k} = - \frac{H^2 \tau}{\sqrt{2}} \sum_\lambda  \frac{1}{k^{1/2}} \vec{\epsilon}_\lambda  \left( \vec{k} \right)  \left[   a_\lambda \left( \vec{k} \right)   +  a_\lambda^\dagger  \left( - \vec{k} \right)  \right] \\ \\
\vec{B}  = \int \frac{d^3 k}{\left( 2 \pi \right)^{3/2}}  \, {\rm e}^{i \vec{k} \cdot \vec{x}} \, \vec{B}_{\vec k} \;\;,\;\;
\vec{B}_{\vec k}  =   \frac{3  H^2 }{\sqrt{2}} \sum_\lambda \lambda  \frac{1}{k^{3/2}} \vec{\epsilon}_\lambda \left( \vec{k} \right)  \left[   a_\lambda \left( \vec{k} \right)   +  a_\lambda^\dagger  \left( - \vec{k} \right)  \right] 
\end{array} \;\;\;,\;\; - k \tau \ll 1
\label{E-B-FT}
\end{eqnarray}
in the super-horizon limit. As for the standard scalar case, the mode function does not oscillate in the super-horizon regime, which is a signal that the field has become classical \cite{Polarski:1995jg}.  The energy densities are given by
\begin{equation}
\rho_E  =  \frac{\langle \vec{E}^2 \rangle}{2} \simeq \frac{ H^4 \, \tau^2}{4 \pi^2} \, \int d k \, k \left( 1 + k^2 \tau^2 \right)\;\;,\;\;
\rho_B  =  \frac{\langle \vec{B}^2 \rangle}{2} \simeq \frac{9 \, H^4 }{4 \pi^2} \, \int \frac{d k}{k} 
\end{equation}
We must only compute the energy in the classical fields, and we therefore limit the integrals to momenta that have exited the horizon during inflation. Modes of smaller wavelength remain in their vacuum state and their contribution to the energy density must be renormalized away (this is part of the cosmological constant problem). The smallest momentum $k_{\rm min} \simeq \frac{1}{-\tau_{\rm in}}$ corresponds to modes that have exited the horizon at the start of inflation (we stress that we are computing  theoretical expectations under a constrained and finite value for the total number  of e-foldings of inflation). For any moment $\tau$, the largest momentum $k_{\rm max} \simeq \frac{1}{-\tau}$ corresponds to modes that have just exited the horizon at that moment. Therefore, for any $\tau \gg \tau_{\rm in}$ during inflation,
\begin{equation}
\rho_E \simeq \frac{3 \, H^4}{16 \pi^2}  \;\;,\;\;
\rho_B \simeq \frac{ 9 \, H^4}{4 \pi^2} \ln \frac{a \left( \tau \right)}{a \left( \tau_{\rm in} \right)} 
\label{rho-EB}
\end{equation}

These behaviors are very different from the usual behavior $\rho \propto a^{-4}$ for radiation.  This shows that energy is being transferred from the inflaton to the gauge field through the $I^2 F^2$ coupling. We also note that the energy in the ``magnetic'' component is greater than that in the ``electric'' one. The energy in the ``magnetic'' component is scale invariant, and the logarithmic increase in the final result is due to the increase of the phase space of the modes that have become classical. We also note, that, despite for most of the super-horizon modes the density $\frac{d \rho_E}{d k}$ is several orders of magnitude smaller than  $\frac{d \rho_B}{d k}$, and decreases with time, the total value of $\rho_E$ is ``only''  logarithmically suppressed with respect to that of $\rho_B$. This is due to the fact that the integral for $\rho_E$  has most of its support in the UV region, where the ``electric'' and ``magnetic'' energy densities are not too different from each other.

In passing, notice that our choice to produce scale-invariant ``magnetic'' fields -- as opposed to ``electric'' fields -- is essentially arbitrary from the perspective of primordial nongaussianity.  Indeed, there is an electric/magnetic duality that leaves the Maxwell equations invariant under the replacement $\vec{E} \rightarrow -\vec{B}$, $\vec{B}\rightarrow \vec{E}$ and $I\rightarrow 1/I$; see \cite{Giovannini:2009xa} for more discussion.  In this case at hand, this means that we can interchange the ``electric'' and ``magnetic'' spectra simply by taking $n=-2$ rather than $n=2$.  The feed-back of these produced fluctuations on the scalar inflaton is essentially unchanged under such a replacement.

\subsection{Backreaction Bounds}

Throughout the discussion above, we have assumed that the produced gauge field fluctuations have a negligible effect on the homogeneous background dynamics during inflation.  To ensure that this assumption is consistent, we must verify several backreaction constraints.  First, we require that the energy density in the ``magnetic'' field is much smaller than the potential energy driving inflation.  Hence, we require 
\begin{equation}
\frac{\rho_B}{V} \simeq \frac{3}{4 \pi^2} \, \frac{H^2}{M_p^2} \,  \ln \frac{a \left( \tau \right)}{a \left( \tau_{\rm in} \right)} \ll 1
\label{B-vs-V}
\end{equation}

A second constraint arises because the produced gauge field fluctuations modify the homogeneous Klein-Gordon equation for the inflaton condensate $\varphi_0(t)$.  To ensure that the usual slow roll equations (\ref{SReqns}) are reliable, we require that the ``driving force'' in  the inflaton equation of  motion  is dominated by the derivative of the inflaton potential.  That is, we require:
\begin{equation}
\label{back2}
 \vert V_{,\varphi} \vert  \gg \left|  \frac{I_{,\varphi}}{I} \langle B^2 \rangle \right|
\end{equation}

Finally, we note that the backreaction of produced gauge fields can also lead to a correction for the effective mass of the inflaton.  The easiest way to see this effect is to note that the action (\ref{model1}) contains a term of the form $(I^2)_{,\varphi\varphi} (\delta\varphi)^2 \langle F^2 \rangle$ once we expand $\varphi = \varphi_0 + \delta\varphi$ and replace $F^2$ which its vacuum average, to estimate the magnitude of backreaction effects.  We require that this ``new'' correction to the inflaton mass is much smaller than the Hubble scale so that we do not spoil the scale invariance of the spectrum.  This amounts to a constraint
\begin{equation}
\label{back3}
  \frac{1}{\epsilon M_p^2} \langle B^2 \rangle \ll H^2
\end{equation}

The conditions (\ref{back2}) and (\ref{back3}) are more stringent than (\ref{B-vs-V}).  In both cases, we obtain a constraint on the total number of e-foldings of inflation of the form
\begin{equation}
\label{cond}
  N_{\mathrm{tot}} \ll 10^{-1} \mathcal{P}^{-1}
\end{equation}
where $N_{\mathrm{tot}} = \ln\left[ a_{\mathrm{end}} / a_{\mathrm{in}}\right]$ and we have defined
\begin{equation}
{\cal P} \simeq \frac{H^4}{4 \pi^2 \dot{\varphi}^2} \simeq \frac{H^2}{8 \pi^2 M_p^2 \epsilon} \simeq 
2.5 \cdot 10^{-9}
\label{power-std}
\end{equation}
which gives the amplitude of the power spectrum from the usual vacuum fluctuations \cite{Baumann:2008aq}.  (In this model there is also an additional contribution to the power spectrum coming from second order effects, however, we will always work in a regime where this is subdominant.  More on this later.)

The condition (\ref{cond}) then indicates that backreaction is negligible provided that $N_{\rm tot} \ll {\rm O } \left(  10^7 \right)$. Note that $N_{\rm tot}$ enters in this condition because the  ``magnetic'' field energy grows (proportionally to the number of e-folds) during the \emph{whole} duration of inflation.  We stress that $N_{\rm tot}$ may be much greater than the number of  e-folds $N_{\rm CMB} \simeq 50 -  60$ which separates the moment at which the largest CMB scales exited the horizon to the end of inflation.

\subsection{Connection with  Magnetogenesis}
\label{subsec:magnetic}

Magnetic fields have been observed at many scales.  They are present in structures (eg - galaxies, galaxy clusters and high redshift protogalactic structures) with strength $\sim 10^{-6}-10^{-3}\,\mathrm{G}$, and in the low density intergalactic space with strength $\sim 10^{-14}-10^{-17} \, \mathrm{G}$.  See \cite{Grasso:2000wj,Giovannini:2003yn,Widrow:2011hs,Kandus:2010nw} for reviews. The origin of these fields is not well understood.  Although an astrophysical mechanism is not ruled out, the observed homogeneity and large coherence length ($\sim \mathrm{kpc}-\mathrm{Mpc}$) could suggest a primordial origin.

For a standard electromagnetic action, the photon is conformally coupled to a FRW geometry; loosely speaking, the scale factor drops from the action term $\sqrt{-g} F^2$, and the photon remains in its vacuum state. Mechanisms for generation of magnetic field during inflation break the conformal invariance by introducing some extra-term. For instance,  in \cite{Turner:1987bw} couplings to the curvature invariants of the type $R A^2$ and $R_{\mu\nu} A^\mu A^\nu$ were considered. These couplings however break the U(1) invariance associated to the electromagnetic field, and one should worry about the longitudinal photon component that they introduce. It was shown in   \cite{Himmetoglu:2008zp,Himmetoglu:2009qi} that, with the $R^2 A^2$ coupling introduced in  \cite{Turner:1987bw}, the longitudinal photon is a ghost. It is safer to consider models that preserve the U(1) invariance.  Axial couplings $\frac{1}{f}\varphi F\tilde{F}$ have been considered in \cite{Anber:2006xt,Durrer:2010mq,Byrnes:2011aa}.  In such models it is typically difficult to produce a sufficiently large field.  Note that any attempt to generate primordial magnetic fields via an axial coupling must take into account the limits on $f$ due to nongaussianity from inverse decay effects \cite{Barnaby:2010vf,Barnaby:2011vw,Barnaby:2011qe}, which are much more stringent than backreaction bounds.

The model (\ref{model1}) has been studied in connection with primordial magnetogenesis \cite{Ratra:1991bn,Gasperini:1995dh,Martin:2007ue,Subramanian:2009fu,Seery:2008ms,Demozzi:2009fu,Caldwell:2011ra}.  It is indeed tempting to identify  $A_\mu$ with the standard model photon.  If we do so, and we assume that the electromagnetic energy density scales in the standard way $\rho \propto a^{-4}$ from the end of inflation on, we find
\begin{equation}
\left( \frac{d \rho_B}{d \ln k} \right)_{\rm today}^{1/2} \simeq 10^{-15} \, {\rm G} \left( \frac{H}{10^{15} \, {\rm GeV}} \right)^{2/3} \left( \frac{T_{\rm rh}}{10^9 \, {\rm GeV}} \right)^{2/3}
\label{Btoday}
\end{equation}
where we have assumed  matter domination due to the coherent inflaton oscillations until reheating takes place at the temperature $T_{\rm rh}$.  We have also disregarded the current departure from matter domination (this gives a negligible correction to the estimate), and treated the value of $H$ as constant during inflation. We note that a lower reheating temperature results in a smaller value of the magnetic field today \cite{Demozzi:2009fu}; the estimate obtained in \cite{Caldwell:2011ra} assumes that radiation domination starts immediately after inflaton. In this case  the expression (\ref{Btoday}) evaluates to $10^{-10} \, {\rm G}  \left( H / 10^{15} \, {\rm GeV} \right) $.

The problem arising in associating the field $A_\mu$ with the electromagnetic photon  has already been stressed in \cite{Demozzi:2009fu}. The model (\ref{model1}) must be supplemented by the action for the matter fields. The most minimal approach is to assume that $I^2 \left( \varphi \right)$ only enters in the $F^2$ term, so that the relevant terms for the electromagnetic coupling of the (Standard Model) fermions are
\begin{equation}
   \mathcal{L}_{\mathrm{matter}} = - \frac{I^2(\varphi)}{4} F^2 - \bar{\psi} \gamma^\mu (\partial_\mu + i e \, A_\mu) \psi  
\label{matter-action}
\end{equation}
If this is the case, the  ``instantaneous''  electric coupling constant is 
\begin{equation}
e_{\rm physical} \equiv e\,  I^{-1}\left[\varphi_0(t)\right]
\end{equation}
During inflation we have $I\propto a^{n}$ and, consequently, for $n>0$, the electric coupling constant decreases by a huge factor during inflation (we recall that the scale invariant $B$ field is obtained for $n=2$).  Thus, if we take $e_{\mathrm{in}} \lsim \mathcal{O}(1)$  at the start of inflation, then the gauge coupling at the end of inflation will be \emph{extremely} tiny.  Assuming no further evolution of $e_{\rm physical} \left( t \right)$ in the post-inflationary epoch, we clearly cannot identify $A_\mu$ with our photon. Alternatively, if we normalize $I$ such that $e_{\rm physical} $ after inflation coincides with the present value, we necessarily imply that $e_{\rm physical} $ was extremely large all throughout inflation, apart from the very last stages.  Even if there were no real charged particles during inflation, this would lead to strong quantum effects from the vacuum fluctuations of these fields, and to a quantum theory (at the very  least) out of computational control.  This poses serious questions on any result obtained from the model. We stress that this problem is not present if $A_\mu$ is a hidden sector gauge field, since in this case one may assume that its associated physical coupling constant is  $\lsim \mathcal{O}(1)$  at the start of inflation.

We briefly comment on a few (unsuccessful) attempts to solve this problem. Firstly, we note that moving the function $I^2 \left( \varphi \right)$   outside the entire electromagnetic-sector Lagrangian does not affect this issue. Indeed,  multiplying the second term of (\ref{matter-action}) by any factor ${\tilde I}$ 
affects both the fermionic kinetic term ${\bar \psi} \partial_\mu \psi$ and the vertex  ${\bar \psi} A_\mu \psi$; however, the fermionic field enters quadratically in both expressions. After canonical normalization of the fermionic field, the factor ${\tilde I}$  drops out from the physical value of the electric coupling constant. It is also difficult to imagine how one may try to modify the structure of the covariant derivative without spoiling gauge invariance.

Secondly, one may try to arrange for a time evolution of $I$ during reheating in such a way that  $e_{\rm physical}$ is brought from a very tiny value at the end of inflation (so to avoid the strong coupling problem during inflation) to the present value before the onset of Big-Bang Nucleosynthesis (given that only a fractional discrepancy from the current value can be tolerated then  \cite{Cyburt:2004yc,Coc:2006sx}). We stress that this requires a huge change of $I$,  which can be difficult to accomplish without disrupting the result for the magnetic field achieved during inflation.  The comoving energy densities 
\begin{eqnarray}
{\bar \rho}_B &=& \frac{1}{2 \pi^2} \int d k \, k^4 V^2 \equiv \int d k \, {\bar \rho}_{Bk} \nonumber\\
{\bar \rho}_E &=& \frac{1}{2 \pi^2} \int d k \, k^2 \left( V' - \frac{I'}{I} \,  V \right)^2 \equiv \int d k \, {\bar \rho}_{Ek} 
\end{eqnarray}
need to satisfy
\begin{equation}
\frac{d}{d \tau} \left( {\bar \rho}_{Ek} +  {\bar \rho}_{Bk} \right) = - 2 \frac{I'}{I} \left(  
 {\bar \rho}_{Ek} -  {\bar \rho}_{Bk} \right)
\end{equation}
If the electric component in this expression can be neglected, one finds ${\bar \rho}_{Bk} \propto I^{2}$; alternatively, if the  magnetic component can be neglected, one finds ${\bar \rho}_{Ek} \propto I^{-2}$. 
In general, achieving such a large change in $I$ during reheating does not appear feasible.

Thirdly, one may abandon the idea of identifying $A_\mu$ with the electromagnetic field, but 
still generate a large scale value of a hidden sector and weakly coupled  $A_\mu$, and then try to convert it to an electromagnetic field through some coupling. For instance, gauge invariance allows for
\begin{equation}
\Delta {\cal L} = \frac{\chi}{2} F^{\mu \nu} \, {\cal F}_{\mu \nu}^{\rm em} \;\;\; {\rm or } \;\;\;
\Delta {\cal L} = \frac{\chi}{2} \epsilon^{\mu \nu\alpha\beta}F_{\mu \nu} \, {\cal F}_{\alpha \beta}^{\rm em} 
\label{paraphoton}
\end{equation}
The first coupling was originally proposed in     \cite{Holdom:1985ag}, and one can promote $\chi$ from a constant parameter to the expectation value of a scalar field; in the second case, $\chi$ is a pseudo-scalar function. One could imagine  that $\chi$  experiences a quick transition from zero to a nonvanishing value $\chi_*$ at some given time $\tau_*$ after inflation, when $I$ has  set to one (we model the transition with a step function; clearly this approximation will break at very small scales). Solving the equations of motion in vacuum at leading order in $\chi_*$, and requiring continuity of the vector potentials at $\tau_*$, one finds that the turning on of $\chi$ results in a partial conversion of the ``electric'' or of the ``magnetic'' component of $A_\mu$ into our electric field: $\vec{\cal E}^{\rm em} \simeq \chi_* \, \vec{E}$ in the first case in (\ref{paraphoton}), and  $\vec{\cal E}^{\rm em} \simeq \chi_* \, \vec{B}$ in the second case. This solution is obtained in absence of any charged particle. However, as soon as a  plasma is formed, it shuts off the electric field well before it can convert into a magnetic field. If already present at  $\tau_*$, the plasma  would prevent any electric field generation at all.

In conclusion, none of these attempts appears to provide a solution to the strong coupling problem.
We believe that a solution, if at all possible, will require a more radical modification of the model than those mentioned here.

\section{Overview of the Mechanism and Phenomenology}
\label{sec:pheno}

In this Section, we describe how the model impacts the cosmological perturbations. The emphasis is in describing the key points, and in summarizing the results. The actual rigorous computations are performed in  Sections \ref{sec:scalar} and \ref{sec:tensor}.

\subsection{Summarized Discussion of the Mechanism}

In the last Section, we showed that the homogeneous condensate $\varphi_0(t) \equiv \langle \varphi(t,{\vec x}) \rangle$ leads to the production of a scale invariant spectrum of primordial ``magnetic'' fields via the coupling $I^2(\varphi) F^2$.  However, this \emph{same} coupling also provides a channel for the produced gauge field fluctuations to feed back into the perturbations of the inflaton $\delta\varphi(t,{\vec x}) \equiv \varphi(t,{\vec x}) - \varphi_0(t)$.  Heuristically, we can see this effect by looking at the equation of motion for the inflaton perturbations.  This has the general form
\begin{equation}
\label{heuristic_eqn}
 \delta\varphi'' + 2\sH \delta\varphi' - \Lap \delta\varphi + a^2 \mathcal{M}^2 \delta\varphi = a^2 \frac{I_{,\varphi}}{I} \left[\vec{E}^2 - \vec{B}^2\right] + \cdots
\end{equation}
where $\sH \equiv a' / a$, $\mathcal{M}^2 \ll H^2$, $\Lap=\partial_i\partial_i$, and $\cdots$ denotes gravitational interactions and also terms involving more derivatives of $I(\varphi)$, all of which give subdominant corrections to the phenomenology.  We see from (\ref{heuristic_eqn}) that the large scale gauge field fluctuations can source inflaton perturbations.  Schematically, the solution of (\ref{heuristic_eqn}) behaves as
\begin{equation}
 \delta\varphi = \underbrace{\delta\varphi_{\mathrm{vac}}}_{\mathrm{homogeneous}} + \underbrace{\delta\varphi_{\mathrm{sourced}}}_{\mathrm{particular}}
\end{equation}
The homogeneous solution is just the usual quantum vacuum fluctuation amplified by the quasi de Sitter expansion, $\delta\varphi_{\mathrm{vac}} \sim H / (2\pi)$.  On the other hand, we have an additional contribution which is sourced by the gauge field fluctuations.  This new sourced contribution is uncorrelated with the vacuum fluctuations, hence its contribution in the $n$-point correlation functions will add incoherently with the standard results.  Moreover, the source contribution to $\delta\varphi$ is highly nongaussian; it is bilinear in the (nearly) gaussian gauge field fluctuations.  The curvature fluctuation $\zeta = -\frac{H}{\dot{\varphi}_0}\delta\varphi$ therefore will also be characterized by a new contribution that is nongaussian and uncorrelated with the vacuum part.

A completely analogous discussion applies also to the tensor perturbations.  Since the produced gauge field fluctuations carry anisotropic stress/energy, they provide a new (essentially classical) source of gravitational waves which is uncorrelated with the standard vacuum fluctuations.

The underlying physics discussed here is very similar to the inverse decay processes that have been computed in \cite{Barnaby:2010vf,Barnaby:2011vw,Barnaby:2011qe} and also the rescattering effects considered in \cite{Barnaby:2009mc,Barnaby:2009dd,Barnaby:2010ke}.  However, in both of those cases the second order ``sourcing'' of inflaton perturbations occurred near horizon crossing, leading  to a (nearly) equilateral bispectrum.  Here, on the other hand, we have a source term that is most significant on very large scales.  This large scale entropy mode leads to a (logarithmic) time evolution of $\zeta_{\mathrm{sourced}}$ on super-horizon scales, and consequently the bispectrum is of nearly local shape.

\subsection{Summary of the Key Phenomenology}

We define the power spectrum of curvature fluctuations via
\begin{equation}
 \langle \zeta_{\vec k} \zeta_{\vec k'} \rangle = \frac{2\pi^2}{k^3} P_\zeta(k) \delta^{(3)}\left({\vec k} + {\vec k'}\right)
\end{equation}
The final result for $P_\zeta$, evaluated on large scales and at the end of inflation, is
\begin{equation}
P_\zeta = {\cal P} \left[ 1 + 192 \, {\cal P} \, N_{\rm CMB}^2 \, \left( N_{\rm tot} - N_{\rm CMB} \right) \right]
\label{power-result}
\end{equation}
where ${\cal P}^{1/2} \equiv \frac{H^2}{2\pi|\dot{\varphi}_0|}$ is the amplitude of the power spectrum of the vacuum modes, $N_{\rm CMB}$ denotes the number of e-folds between the moment at which the large scales CMB modes leave the horizon and the end of inflation, and $N_{\rm tot}$ is the total number of e-folds of inflation.  The first term in (\ref{power-result}) is the usual contribution from the vacuum while the second term is the ``sourced'' contribution described heuristically in the last Subsection.  Here we work to leading order in slow roll parameters, so the spectrum (\ref{power-result}) is exactly flat.  In a more complete computation we would see small departures from scale invariance $\propto k^{n_s-1}$.

We can physically understand the structure of (\ref{power-result}) as follows.  The suppression $\mathcal{P}\ll 1$ in the second term arises simply because we are computing an effect which is higher order in perturbation theory.\footnote{See \cite{Barnaby:2011pe} for more discussion on the counting of such factors in models with particle production.}  The factors of $N_{\mathrm{CMB}}$, on the other hand, arise due to the logarithmic time evolution, $\zeta_{\mathrm{sourced}} \sim \ln a$, outside the horizon.  Such growth is consistent since we have large scale entropy perturbations playing an important role in the dynamics.  Finally, the factor $N_{\mathrm{tot}} - N_{\mathrm{CMB}}$ is related to the phase space of contributing gauge field fluctuations.  It is related to the number of $B$-modes that source the inflaton perturbation, and it is the counterpart of the logarithmic enhancement in the background density (\ref{rho-EB}). We explain this in details after eq. (\ref{dfdf-res}).   We should stress that equation (\ref{power-result}) is valid only when $N_{\rm tot} \gg N_{\rm CMB}$.  Otherwise the factor $N_{\rm tot} - N_{\rm CMB} $ is replaced by an order one factor.

Throughout this paper we will require that the sourced contribution to the power spectrum is subdominant:
\begin{equation}
\label{subdominant-sourced}
 192 \, {\cal P} \, N_{\rm CMB}^2 \, \left( N_{\rm tot} - N_{\rm CMB} \right) < 1
\end{equation}
This yields a constraint on the total number of e-foldings $N_{\mathrm{tot}} - N_{\mathrm{CMB}} < \mathcal{O}(10^{-6}) \mathcal{P}^{-1}$ (taking $N_{\mathrm{CMB}} \sim 60$ for illustration) which is considerably more stringent than the backreaction bounds discussed in the last Section.  

The bispectrum is given by the 3-point correlation function
\begin{equation}
 \langle \zeta_{\vec k_1}\zeta_{\vec k_2} \zeta_{\vec k_3} \rangle \equiv B_\zeta \left( k_i \right) \,
 \delta^{(3)}\left({\vec k}_1 + {\vec k}_2 + {\vec k}_3 \right)
\end{equation}
We have found that our bispectrum is very close to the local shape.  Indeed, the ``cosine'' between our bispectrum and the local shape, as defined in \cite{shape} (this is a measure on how well a template reproduced a given bi-spectrum), is about $0.98$.  Nevertheless, here we will retain the full momentum dependence of the bispectrum, since it has a simple analytical shape. 

As is conventional in the literature (see e.g. \cite{Komatsu:2001rj}), we defined a $k$-dependent nonlinearity parameter from computing the bispectrum and the power spectrum of $\zeta$, and by comparing them with those obtained from
\begin{equation}
\zeta \left( {\vec x} \right) = \zeta_g \left( {\vec x} \right) + \frac{3}{5} \, f_{\rm NL} \,\left[ \zeta_g^2 \left( {\vec x} \right) - \langle  \zeta_g^2 \left( {\vec x} \right) \rangle \right]
\end{equation}
where $\zeta_g$ is gaussian. An  explicit computation gives the result
\begin{eqnarray}
f_{NL}(k_i) &\simeq& f_{NL}^{\rm equiv. \;  local} \times \frac{3}{4} \,  \frac{ \frac{1+\cos^2 \left( \vec{k}_1 ,\, \vec{k}_2 \right)}{ k_1^3 \, k_2^3} +  \frac{1+\cos^2 \left( \vec{k}_1 ,\, \vec{k}_3 \right)}{ k_1^3 \, k_3^3}  +  \frac{1+\cos^2 \left( \vec{k}_2 ,\, \vec{k}_3 \right)}{ k_2^3 \, k_3^3} }{\frac{1}{k_1^3 \, k_2^3} + \frac{1}{k_1^3 \, k_3^3} + \frac{1}{k_2^3 \, k_3^3} } \nonumber\\
 f_{NL}^{\rm equiv. \;  local}  &\simeq&  1280 \, \mathcal{P} N_{\rm CMB}^3 \left( N_{\rm tot} - N_{\rm CMB} \right) \nonumber\\
&\simeq&   0.7  \left( \frac{N_{\rm CMB}}{60} \right)^3 \left( N_{\rm tot} - N_{\rm CMB} \right)
  \label{fNL-summary}
\end{eqnarray}
where we have assumed that (\ref{subdominant-sourced}) is satisfied. We can a-posteriori see that this condition is indeed  always satisfied whenever the result (\ref{fNL-summary}) is within the observational limits.   

We have defined our ``equivalent'' local nonlinearity parameter as follows:  we first note that both our bispectrum and the local template are enhanced in the squeezed limit. In this limit,  our bispectrum satisfies
\begin{eqnarray}
k_1^3 \, k_3^3 \, \langle \zeta_{\vec{k}_1} \,  \zeta_{\vec{k}_2} \,  \zeta_{\vec{k}_3}
 \rangle & \propto & 1 + \cos^2 \left( \vec{k}_1 ,\, \vec{k}_3 \right)
\;\;\;\;\;\;,\;\;\;\;\;\;  k_3 \ll k_1 \simeq k_2 \nonumber\\
&\propto& \cos \epsilon \; Y_0^0 + \sin \epsilon \,  Y_2^0
\;\;\;\;\;\;,\;\;\;\;\;\;  \epsilon \equiv \tan^{-1} \frac{1}{2 \sqrt{5}} \simeq 0.22
\label{Ylm-squeezed}
\end{eqnarray}
where $Y_l^0$ are normalized spherical harmonics, characterized by the angle between $\vec{k}_1$ and $\vec{k}_3$.  The average of  $f_{NL}$ over all possible values of this angle is then equal to the average of $f_{NL}^{\rm equiv. \;  local}$.  Note that the current CMB  limit on local shape nongaussianity is $-10 < f_{NL}^{\mathrm{local}} < 74$ at $95\%$ CL \cite{Komatsu:2010fb}.

Using the local template to study this signature is clearly a good approximation, given that the local template is characterized by the monopole part only in (\ref{Ylm-squeezed}), and that $\cos \, \epsilon \simeq 0.98$. However, we note more than $1/5$ of the amplitude in (\ref{Ylm-squeezed}) is contributed by the quadrupole part. This is in contrast to what typically happens in non-gaussianity from scalar fields only, where the quadrupole and higher harmonic terms in the squeezed limit expansion are suppressed by powers of $k_3/k_1$ and give a negligible contribution in this limit \cite{Lewis:2011au}.
The angular dependence is imprinted by a ``directionality'' generated by the largest wavelength mode $k_3^{-1}$, seen by the smaller modes when they leave the horizon. In the scalar case, the directionality is typically due to a gradient, and therefore it vanishes in the $k_3 \rightarrow 0$ limit. In our case, the directionality is due to the polarization of the vectors, and it therefore remains finite in the limit. 

Let us also discuss the parametrical dependence of  $f_{NL}^{\rm equiv. \;  local}$. The factor $N_{\rm CMB}^3$ is due to the super-horizon growth of the three modes used in computing the three point function. The factor $N_{\rm tot} - N_{\rm CMB}$ is due to the number of super-horizon modes that contribute to the correlator, analogously to what we have described in relation to (\ref{power-result}).  Also in this case, this factor is replaced by an order one factor if $N_{\rm tot}$ is close to $N_{\rm CMB}$.

The ``magnetic'' fields also produce gravity wave modes, which add incoherently with the vacuum ones. The power of gravity waves produced during inflation is conventionally parametrized by the ratio $r$  of their power divided by the scalar power. We find
\begin{equation}
r \equiv \frac{P_{\rm GW}}{P_\zeta} \simeq 16 \, \epsilon \,  \frac{1+48 \, \epsilon \, {\cal P} \, N_{\rm CMB}^2 \left( N_{\rm tot} - N_{\rm CMB} \right)}{1+192   \, {\cal P} \, N_{\rm CMB}^2 \left( N_{\rm tot} - N_{\rm CMB} \right)}
\end{equation}
which gives the standard result $r\approx 16 \epsilon$ when the vacuum modes dominate the power spectrum of curvature fluctuations.

\section{Scalar Perturbations}
\label{sec:scalar}

To encode the effect of the gauge field on the cosmological perturbations we need to study the perturbations up to second order. Therefore, we decompose
\begin{equation}
\varphi = \varphi_{0}   + \delta_1 \varphi + \delta_2 \varphi \;\;,\;\;
g_{\mu \nu} = g_{\mu \nu \,0}   + \delta_1 g_{\mu \nu}  + \delta_2 g_{\mu \nu} 
 \;\;,\;\; A_\mu
\end{equation}
The gauge field has no zero order part, and, as we discuss below, we do not need to evaluate it at second order.

Since the gauge field has no zero order part, the metric/inflaton perturbations do not mix with the gauge field modes at linear order; this is because the gauge field enters already quadratically in the action for the perturbations, through the expansion of the last term in (\ref{model1}). This is the same reason that in the previous Section allowed us to compute $A_\mu$ disregarding inflaton and metric perturbations. We note that the gauge field can still affect the first order metric/inflaton perturbations through its backreaction on the background evolution. This can be disregarded under the assumption that the two conditions (\ref{B-vs-V}) and (\ref{cond}) hold.

Therefore, at the linearized level the standard results of scalar field inflation hold. We work in the spatially flat gauge for the scalar perturbations, $\delta g_{ij} =0$. In this gauge, the curvature perturbation on uniform density hypersurfaces is $\zeta = - \frac{H}{\dot{\varphi}_0} \, \delta \varphi$.
As we show in Appendix \ref{appA}, one finds
\begin{equation}
\label{eq-d1}
 \left[ \frac{\partial^2}{\partial\tau^2} + 2\sH \frac{\partial}{\partial\tau} - \Lap + \left( a^2 V_{,\varphi\varphi}
 - 3\frac{(\varphi_0')^2}{M_p^2}  \right) \right] \delta_1 \varphi = 0 \, ,
\end{equation}
which is the standard equation for the Mukhanov-Sasaki \cite{mukhanov,sasaki} variable rewritten in terms of $\zeta$; the last term in this equation has been simplified using slow roll approximation. Finally, ${\cal H} = \frac{a'}{a} = a H$.  Here we have disregarded corrections to the effective mass due to backreaction effects; see Section \ref{sec:EB}.  In the end we will only perform computations at leading order in slow roll parameters, hence this neglect has no impact on our final results.

To compute the effect of the gauge fields on the cosmological perturbations we expand all the equations of the system at second order in the perturbations. We combine these equations in the same formal way that they are combined to obtain  (\ref{eq-d1}). In this way we obtain a ``master equation'' for $\delta_2 \varphi$ that does not contain any $\delta_2 g_{\mu \nu}$ mode. As we show in Appendix  \ref{appA}, this equation reads
\begin{equation}
 \left[ \frac{\partial^2}{\partial\tau^2} + 2\sH \frac{\partial}{\partial\tau} - \Lap + \left( a^2 V_{,\varphi\varphi} - 3\frac{(\varphi_0')^2}{M_p^2}  \right) \right] \delta_2 \varphi 
 = J_1 \left[ A^2 \right] + J_2 \left[  \left(  \delta_1 \varphi \right)^2 ,\,  \left(  \delta_1 g \right)^2 ,\,  \delta_1 \varphi \times \delta_1 g \right] 
\label{eq-d2}
\end{equation}
where
\begin{equation}
J_1 \left[ A^2 \right] \equiv \frac{a^2}{2} \frac{I^2_{,\varphi}}{I^2} \left(\vec{E}^2 - \vec{B}^2\right)  - \frac{a^2 \varphi'_{0}}{2\sH M_p^2} \left[ \frac{\vec{E}^2 + \vec{B}^2}{2}   + \frac{1}{a^4} \Lap^{-1}\partial_\tau\left(a^4 \vec{\nabla}\cdot (\vec{E}\times\vec{B})\right)   \right] 
\label{Jem}
\end{equation}
As for the linear theory equation, we have disregarded a backreaction-induced correction to the effective mass.

Note that our model (\ref{model1}) contains higher dimension interactions between the inflaton and gauge fields of the form $\sum_n c_n (\delta\varphi)^n F^2$ which arise from expanding the coupling function $I^2(\varphi)$ in powers of $\delta\varphi$.  These couplings will enter into the calculation explicitly at higher order in perturbation theory.  Using the in-in formalism, we have verified that such high dimension operators do not modify our leading order results for the spectrum and bispectrum. 

The right hand side of (\ref{eq-d2}) comprises of two sources for $\delta_2 \varphi$; the first source 
contains terms at second order in the gauge perturbations, and has been completely given 
in (\ref{Jem}). The second source contains terms that are the product of two first order inflaton perturbations, or of two first order metric perturbations (not only the scalar ones), or of one first order inflaton perturbations times one first order metric perturbation. We note that no ``mix source'' of the type $\delta_1 \varphi \times A$ or  of the type $\delta_1 g_{\mu \nu} \times A_\mu$ in present in  (\ref{eq-d2}), because $A_\mu$ does not enter linearly in (\ref{model1}). 

Expression (\ref{eq-d2})-(\ref{Jem}) was first obtained in \cite{Seery:2008ms}  by extremizing the cubic order action of the perturbations. This is equivalent to working directly with the equations expanded at second order, and we have verified that our result coincides with that of  \cite{Seery:2008ms}. The source $J_2$ is the standard result obtained at second order in single scalar field inflation. The scalar part of this expression in the gauge we have adopted is explicitly given in  \cite{Malik:2006ir}.

Therefore, the inflaton perturbation is formally given by
\begin{equation}
\delta \varphi = \left(  \delta_1 \varphi + \delta_2 \varphi \vert_{{\rm sourced \; by \; } J_2} \right)
+  \delta_2 \varphi \vert_{{\rm sourced \; by \; } J_1}  
\label{df-formal}
\end{equation}
The part in parenthesis is the standard result obtained in single scalar field inflation, with only negligible corrections coming from the backreaction of the gauge field on the background dynamics. 
The $ \delta_2 \varphi \vert_{{\rm sourced \; by \; } J_2} $ term is clearly negligible in the primordial power spectrum, and also leads to unobservable non-gaussianity \cite{Acquaviva:2002ud,Maldacena:2002vr,Seery:2005wm,Seery:2008qj}.  We therefore disregard it in this work. The last term in (\ref{df-formal}) encodes the effect of the gauge field on the inflaton perturbations. We note that this term is uncorrelated with the other two, since the quantum/statistical operators entering in $J_1$ are those of the gauge field. Therefore, we are interested in computing
\begin{equation}
\left\langle \delta \varphi^2 \right\rangle  \simeq  \left\langle \left( \delta \varphi_{\rm vacuum} \right)^2 \right\rangle  +  \left\langle \left( \delta \varphi_{\rm sourced} \right)^2 \right\rangle  
\;\;\;,\;\;\; 
\left\langle \delta \varphi^3 \right\rangle  \simeq    \left\langle \left( \delta \varphi_{\rm sourced} \right)^3 \right\rangle  
\end{equation}
where
\begin{equation}
\delta  \varphi_{\rm vacuum}  \equiv \delta_1 \varphi \;\;\;,\;\;\;
\delta  \varphi_{\rm sourced}  \equiv \delta_2 \varphi  \vert_{{\rm sourced \; by \; } J_1}  
\label{def-df}
 \end{equation}

We combine (\ref{eq-d1}) and (\ref{eq-d2}) in a unique equation for $\delta \varphi = \delta \varphi_{\rm vacuum} + \delta \varphi_{\rm sourced}$, where the two quantities are defined in (\ref{def-df}). We approximate this equation in slow roll approximation and we keep only the leading source term that arises from the direct $I^2 F^2$ coupling. This gives
\begin{equation}
\left[ \partial_\tau^2 + 2 \frac{a'}{a} \partial_\tau - \Lap \right] \delta \varphi \simeq J \;\;\;,\;\;\;
J = \frac{a^2}{2} \, \frac{I^2_{,\varphi}}{I^2} \left[ E^2 - B^2 \right]
\simeq - \sqrt{\frac{2}{\epsilon}} \, \frac{a^2}{M_p} \,  \left[ E^2 - B^2 \right]
\label{final-eq-df}
\end{equation}

We note that the source $E^2-B^2$ interacts with the inflaton perturbation with a strength that is gravitationally suppressed but slow roll $1/\sqrt{\epsilon}$ enhanced (this is one of the enhancements that make non-gaussianity visible in the model). The remaining terms in (\ref{Jem}) have the same scale dependence, but an interaction strength that is both gravitationally and slow roll $\sqrt{\epsilon}$ suppressed. The same  suppression characterizes all the terms in $J_2$. Therefore, the dominant source is that one arising from the direct inflaton-gauge field coupling $I^2 \left( \varphi \right) F^2$.  We  explicitly see that simply computing the effect on $\delta \varphi$ from the gauge fields in an unperturbed metric reproduces the leading results for $\delta_2 \varphi$.

In this calculation, we have estimated the curvature perturbation in flat gauge as $\zeta = -\frac{\sH}{\varphi_0'}\delta\varphi$. This equation actually receives corrections at second order; however, these corrections are (1) subdominant at the end of inflation, and (2) become even smaller (by several orders of magnitude) during reheating, when the energy in the gauge field decreases faster than the one of the dominating plasma (the equation of state of the dominating plasma is either the one of matter - for perturbative reheating -  or intermediate between the one of matter and radiation \cite{Podolsky:2005bw} - in the nonperturbative case). We explicitly show this in Appendix \ref{app:zeta}.

\subsection{The two point correlation function, and the correction to the power spectrum}

We are interested in the primordial curvature perturbation, given by
\begin{equation}
  \zeta(t,{\vec x}) = -\frac{H}{\dot{\varphi}_0} \delta\varphi(t,{\vec x})
\end{equation}
The two point correlation function in momentum space is related to the power spectrum by the standard expression
\begin{equation}
\frac{H^2}{\dot{\varphi}_0^2} \, \left\langle \frac{\delta \varphi_{\vec k}}{a} \, 
\frac{ \delta \varphi_{{\vec k}'}}{a} \right \rangle  =  \left \langle \zeta_{\vec k} \zeta_{{\vec k}'}  \right \rangle \equiv  P_\zeta \left( k \right) \, \frac{2\pi^2}{k^3} \delta^{(3)} \left( \vec{k} + \vec{k}' \right)
 \label{power}
\end{equation}
where all quantities are evaluated at some time $\tau$, and where we use the convention
\begin{equation}
  \delta \varphi (\tau,{\vec x}) = \frac{1}{a}  \int \frac{d^3k}{(2\pi)^{3/2}} \, \delta \varphi_{\vec k}(\tau) \,  e^{i {\vec k}\cdot {\vec x}}
\;\;\;,\;\;\;
\delta \varphi_{\vec k} = \delta \phi_k \, a \left( \vec{k} \right) +  \delta \phi_k^* \, a^\dagger \left( - \vec{k} \right)  
\label{dphi-F}
\end{equation}
We define the Fourier transform of the source as
\begin{eqnarray}
J_{\vec{k}} \equiv a \int \frac{d^3 x}{\left( 2 \pi \right)^{3/2}} \, {\rm e}^{-i \vec{k} \cdot \vec{x}} J 
=  - \sqrt{\frac{2}{\epsilon}} \, \frac{a^3}{M_p} \int \frac{d^3 p}{\left( 2 \pi \right)^{3/2}} \left[ \vec{E}_{\vec p} 
\cdot \vec{E}_{{\vec k} - \vec{p}} -  \vec{B}_{\vec p} \cdot \vec{B}_{{\vec k} - \vec{p}} \right]
\label{J-F}
\end{eqnarray}
where (\ref{E-B-FT}) and (\ref{final-eq-df})  have been used. We find
\begin{eqnarray}
J_{\vec k} \left( \tau \right) \simeq - \frac{H^4 \, a^3 \left( \tau \right) }{ \sqrt{2 \epsilon} \, M_p} \sum_{\lambda ,\lambda'} && \int \frac{d^3 p}{\left( 2 \pi \right)^{3/2}} \, \vec{\epsilon}_\lambda \left( \vec{p} \right) \cdot \vec{\epsilon}_{\lambda'} \left( \vec{k} - \vec{p} \right) \left[ \frac{\tau^2}{\vert \vec{p} \vert^{1/2} \, \vert \vec{k} - \vec{p} \vert^{1/2}} - \frac{9 \lambda \lambda'}{\vert \vec{p} \vert^{3/2} \, \vert \vec{k} -   \vec{p} \vert^{3/2}} \right] \nonumber\\ 
&& \times \left[ a_\lambda \left( \vec{p} \right) + a_\lambda^\dagger \left( - \vec{p} \right) \right] \, 
  \left[ a_{\lambda'} \left( \vec{k} - \vec{p}  \right) + a_{\lambda'}^\dagger \left( - \vec{k} + \vec{p}  \right) \right]  \nonumber\\
\label{source}
\end{eqnarray}
We stress that the first term in the square parenthesis in the first line of (\ref{source}) is the $E$-contribution to the source, while the second term is the $B-$contribution.

Combining eqs. (\ref{final-eq-df}), (\ref{dphi-F}), and (\ref{J-F}), the equation for the Fourier modes of the inflaton perturbations reads
\begin{equation}
\delta \varphi_{\vec k}'' + \left( k^2 - \frac{a''}{a} \right) \delta \varphi_{\vec k} \simeq J_{\vec k}
\label{eq-df}
\end{equation}
where all quantities are evaluated at the time $\tau$.

The homogeneous solution to this equation is the standard vacuum solution, which leads to the standard result (\ref{power-std}) for the power spectrum (\ref{power}). The homogeneous equation is obtained with the Green function method
\begin{equation}
\delta \varphi_{\vec k} \vert_{\rm sourced} = \int_{\tau_{\rm in}}^\tau d \tau' G_{k} \left( \tau ,\, \tau' \right) J_{\vec k} \left( \tau' \right)
\label{formal-sol}
\end{equation}
The leading order result in slow roll approximation is obtained by using the   retarded Green function
of eq. (\ref{eq-df}) in  de Sitter space.
\begin{eqnarray}
G_k \left( \tau ,\, \tau' \right) & \simeq & \frac{1}{k^3 \, \tau \, \tau'} \left[ k \tau' \, \cos \left( k \tau ' \right) - \sin \left( k \tau' \right) \right] \;\;\;,\;\;\; \left| k \, \tau \right| \ll 1 \nonumber\\
& \simeq & - \frac{\tau'^{2}}{3 \, \tau} \;\;\;,\;\;\;  \left| k \, \tau \right| ,\, \left| k \, \tau' \right| \ll 1   \label{GreenFunction}
\end{eqnarray}

Using the source (\ref{source}), and the identity 
\begin{equation}
\left| \vec{\epsilon}_\lambda \left( \vec{p} \right) \cdot  \vec{\epsilon}_{\lambda'} \left( \vec{q} \right) \right|^2 
= \frac{1}{4} \left( 1 - \lambda \, \lambda' \, {\hat p} \cdot {\hat q} \right)^2
\end{equation}
(where hat denotes a unit vector) we obtain
\begin{eqnarray}
\left\langle \frac{\delta \varphi_{\vec k}}{a} \,  \frac{\delta \varphi_{{\vec k}'}}{a} \right\rangle_\tau 
\vert_{\rm sourced}  &\simeq&
\frac{H^4 \, \delta^{(3)} \left( \vec{k} + \vec{k}' \right)}{9 \epsilon M_p^2 \, k^3} \int \frac{d^3 q}{\left( 2 \pi \right)^3} \, \int \, \frac{d y'}{y'} \,  \int  \, \frac{d y''}{y''}   
\nonumber\\
&&\quad\times \left\{ \left[  1 + \cos^2 \left( \vec{q} ,\, \hat{k} - \vec{q} \right) \right] \, \left[ \frac{y'^2 \, y''^2}{\left| \vec{q} \right| \, \left| \hat{k} - \vec{q} \right|} + \frac{81}{\left| \vec{q} \right|^3 \, \left| \hat{k} - \vec{q} \right|^3} \right] + 18  \cos \left( \vec{q} ,\, \hat{k} - \vec{q} \right) \, \frac{y'^2+y''^2}{\left| \vec{q} \right|^2 \, \left| \hat{k} - \vec{q} \right|^2}  \right\} \nonumber\\
\label{EB-dphi}
\end{eqnarray}
where we have introduced the dimensionless integration variables $\vec{q} = \vec{p} / k$, $y' = - k \tau'$, and $y'' = - k \tau''$. The momentum integral need to be restricted so that the gauge modes participating in the original convolution were inside the horizon at the start of inflation (otherwise they would not be produced by this mechanism). This means
\begin{equation}
\vert \vec{q} \vert \;,\; \vert {\hat k} - {\vec q} \vert > \frac{1}{k \, \vert \tau_{\rm in} \vert}
\label{condition2}
\end{equation}
The time integrations are instead restricted to times which are between $\tau_{\rm in}$ and $\tau$, and for which the sourcing modes have exited the horizon. This means $ \tau_{\rm in} ,\, -  \frac{1}{\vert \vec{p} \vert} ,\, - \frac{1}{\vert \vec{k} - \vec{p} \vert} < \tau' ,\, \tau'' < \tau$. We do not need to include $\tau_{\rm in}$ in this condition thanks to (\ref{condition2}). Therefore, the time integrals in (\ref{EB-dphi}) are restricted to
\begin{equation}
k \, \vert \tau \vert < y' ,\, y'' < {\rm Min} \left[ \frac{1}{  \vert \vec{q} \vert } ,\, \frac{1}{  \vert {\hat k} - \vec{q} \vert } \right]  
\end{equation}

In Section \ref{sec:EB} we saw that the energy density in the produced  $B-$field is logarithmically enhanced with respect to that in the $E-$field. Eq. (\ref{EB-dphi}) shows that the an analogous logarithmic enhancement takes place in the source of $\delta \varphi$. In this case, all the three integrals in  (\ref{EB-dphi}) present an enhancement. Disregarding the subdominant $E-$contribution, and performing the time integrals, the expression (\ref{EB-dphi}) gives
\begin{equation}
\left\langle \frac{\delta \varphi_{\vec k}}{a} \,  \frac{\delta \varphi_{{\vec k}'}}{a} \right\rangle_\tau 
\vert_{\rm sourced}  \simeq
\frac{9 H^4}{\epsilon M_p^2} \, \frac{\delta^{(3)} \left( \vec{k} + \vec{k}' \right)}{k^3}  \,  \int \frac{d^3 q}{\left( 2 \pi \right)^3} \, \frac{ 1 + \cos^2 \left( \vec{q} ,\, \hat{k} - \vec{q} \right) }{ 
\left| \vec{q} \right|^3 \, \left| \hat{k} - \vec{q} \right|^3}
\ln^2 \frac{ {\rm Min} \left[ \frac{1}{  \vert \vec{q} \vert } ,\, \frac{1}{  \vert {\hat k} - \vec{q} \vert } \right]    }{k \vert \tau \vert}
\end{equation}
The momentum integral has most of its support at the two logarithmic poles; due to the symmetry between the two poles, we can simply evaluate the integral for $\vert \vec{q} \vert \simeq \frac{1}{k \vert \tau_{\rm in} \vert} \ll 1$ and multiply the result by two:
\begin{equation}
\left\langle \frac{\delta \varphi_{\vec k}}{a} \,  \frac{\delta \varphi_{{\vec k}'}}{a} \right\rangle_\tau 
\vert_{\rm sourced}  \simeq \frac{12}{\pi^2} \, \frac{H^4}{\epsilon \, M_p^2}  \, \frac{\delta^{(3)} \left( \vec{k} + \vec{k}' \right)}{k^3} \, \ln^2 \frac{1}{k \vert \tau \vert} \,  \ln \left( k \vert \tau_{\rm in} \vert \right)
\label{dfdf-res}
\end{equation}
The first logarithmic enhancement is due to the growth of the two modes $ \frac{\delta \varphi_{\vec k}}{a} $ and $ \frac{\delta \varphi_{{\vec k}'}}{a} $ in the super-horizon regime. The growth is due to the presence of the entropy modes $A_\mu$. At the end of inflation $ \ln^2 \frac{1}{k \vert \tau \vert} = N_{\rm CMB}^2$. The second enhancement $ \ln \left( k \vert \tau_{\rm in} \vert \right) = N_{\rm tot} - N_{\rm CMB}$ is due to the number of gauge field modes  that source the inflaton perturbation. More specifically, we have seen that each sourced inflaton mode is obtained as a convolution of two gauge field modes. The momenta of these two modes need to add up to the momentum of the inflaton mode. The second enhancement occurs in the IR limit   of one of the two gauge modes. This enhancement is the counterpart of the enhancement taking place for $\rho_B$, which is also due to the number of large wavelength modes produced during inflation.

As we discussed, the contribution (\ref{dfdf-res}) adds up incoherently with the vacuum one in (\ref{power}). Using the relation $\zeta = -\frac{H}{\dot{\varphi}_0}\delta\varphi$, the sum gives
\begin{equation}
P_\zeta \vert_{\rm end \; inflation} \simeq {\cal P} \left[ 1 + 192 \, {\cal P} \, 
N_{\rm CMB}^2 \, \left( N_{\rm tot} - N_{\rm CMB} \right) \right]
\label{res-power}
\end{equation}
where we recall that $ {\cal P} $ is the contribution from the vacuum term, for which the standard slow roll expression (\ref{power-std}) holds. We also remind that $N_{\rm CMB} \simeq 50-60$ is the number of e-folds before the end of inflation when the largest scale CMB modes left the horizon, while $N_{\rm tot}$ is the total number of e-folds of inflation. The  enhancement from the momentum integral takes place for $N_{\rm tot} \gg N_{\rm CMB}$; if inflation only lasted about the observed number of e-folds, then we estimate that the final  momentum integral produces an order one result, so that the result (\ref{res-power}) remains valid as an order of magnitude estimate.

For $N_{\rm tot} \gg N_{\rm CMB} \sim 60$, the ratio between the sourced and the standard power spectrum is $\simeq 1.7 \cdot 10^{-3} \, N_{\rm tot}$. The standard term dominates provided that inflation lasted less than about $600$ e-folds. In the work, we assume that this is the case.

\subsection{The three point correlation function, and observable non-gaussianity}

We are interested in the three point correlation function of $\zeta$ as a measure of non-gaussianity.
A common parametrization of nongaussianity is the nonlinearity parameter $f_{NL}$, introduced by assuming that the curvature perturbation may be expanded as
\begin{equation}
\label{local_first}
\zeta \left( {\vec x} \right) = \zeta_g \left( {\vec x} \right) + \frac{3}{5} \, f_{\rm NL} \,\left[ \zeta_g^2 \left( {\vec x} \right) - \langle  \zeta_g^2 \left( {\vec x} \right) \rangle \right]
\end{equation}
where $\zeta_g(x)$ is a gaussian random field (see our discussion in \cite{Barnaby:2011vw} for a detailed explanation of the sign conventions  and $2 \pi$ factors in the following expressions).  Both $\zeta$ and $\zeta_g$ may be decomposed as in (\ref{dphi-F}) so that the relation between the q-modes of the Fourier decomposition is
\begin{equation}
\zeta_{\vec k} = \zeta_{g,{\vec k}} + \frac{3}{5} \, f_{\rm NL} \, \int \frac{d^3 p}{\left( 2 \pi \right)^{3/2}} \, 
\zeta_{g,{\vec k}} \, \zeta_{g,{\vec k} - {\vec p} } 
\end{equation}
The three point correlator of $\zeta_g$ vanishes, as this field is gaussian. However, due to the quadratic term in (\ref{local_first}), the three point correlator of $\zeta$ is nonvanishing, and can be 
expressed through a sum of two point correlators of $\zeta_g$.  One finds
\begin{equation}
\langle \zeta_{\vec k_1} \,  \zeta_{\vec k_2} \,  \zeta_{\vec k_3} \rangle = \frac{ 3 }{ 10 } \, \left( 2 \pi \right)^{5/2} \, f_{\rm NL} P_\zeta \left( k \right)^2 \delta^{(3)} \left( {\vec k_1} + {\vec k_2}  + {\vec k_3} \right) \, \frac{\sum_i k_i^3}{\Pi_i k_i^3}
\label{local_bi_first}
\end{equation}
where the power spectrum was defined in (\ref{power}). To obtain this expression,  one  identifies the two point function of $\zeta$ with that of $\zeta_g$ (as the difference is subleading in a perturbative expansion), and  disregards the  mild scale dependence of the power spectrum.

By evaluating $\langle \zeta^3 \rangle$, and by inserting it in (\ref{local_bi_first}), one  defines an ``effective'' (momentum dependent) nonlinearity parameter, even when the intrinsic nongaussianity is not of the local form (\ref{local_first}). The dependence of $f_{\rm NL}$ on the relative size of the momenta is denoted as ``shape'' of the non-gaussianity. Ref \cite{shape} provides a method to evaluate whether the shape obtained in a given model is well reproduced by the local template (namely, $f_{\rm NL}$ constant in (\ref{local_bi_first})) or by any other template employed in data analysis.

As nongaussianity from the vacuum term is negligible, we need to compute
\begin{equation}
\left\langle  \zeta_{\vec k_1} \,  \zeta_{\vec k_2} \,  \zeta_{\vec k_3} \right\rangle \simeq 
-\frac{H^3}{\dot{\varphi}_0^3} \, \left\langle \frac{\delta \varphi_{\vec k_1} }{a} \, 
 \frac{\delta \varphi_{\vec k_2} }{a} \,  \frac{\delta \varphi_{\vec k_3} }{a} \, \right\rangle \vert_{\rm sourced}
\end{equation}
where each expression is evaluated at some given time $\tau$.

We proceed as in the previous Subsection by inserting the source (\ref{source}) into (\ref{formal-sol}) and by evaluating the correlator. Keeping only the dominant ``magnetic'' source, we obtain
\begin{eqnarray}
\left\langle \prod_{i=1}^3  \frac{\delta \varphi_{\vec k_i} }{a}  \right\rangle_\tau \vert_{\rm sourced}
&\simeq& 
\frac{1}{a^3 \left( \tau \right)} \, \frac{729}{8 \pi^{9/2}} \, \frac{H^3}{\epsilon^{3/2} M_p^3}
\, \prod_{i=1}^3 
\int d \tau_i \frac{G_{k_i} \left( \tau,\, \tau_i \right)}{\left( - \tau_i \right)^3} \sum_{\lambda_i \, \sigma_i}   \int d^3 p_i \, \frac{
\vec{\epsilon}^{(\lambda_i)} \left( \vec{p}_i \right) \cdot \vec{\epsilon}^{(\sigma_i)} \left( \vec{k}_i -  \vec{p}_i   \right) }{\vert \vec{p}_i \vert^{3/2} \, \vert \vec{k}_i - \vec{p}_i \vert^{3/2}} \nonumber\\
&&    \quad\quad\quad\quad
   \times
 \delta_{\sigma_1 \lambda_2} \delta^{(3)} \left( \vec{k}_1 - \vec{p}_1 + \vec{p}_2 \right)
 \delta_{\sigma_2 \lambda_3} \delta^{(3)} \left( \vec{k}_2 - \vec{p}_2 + \vec{p}_3 \right)
  \delta_{\sigma_3 \lambda_1} \delta^{(3)} \left( \vec{k}_3 - \vec{p}_3 + \vec{p}_1 \right)
\nonumber\\
\end{eqnarray}
(where the $\delta-$functions emerge from commutators between $a$ and $a^\dagger$ gauge field operators in the standard way) where the integration regions are bounded in an analogous way as for the two point function:
\begin{equation}
\vert \vec{p}_i \vert ,\; \vert \vec{k}_i - \vec{p}_i \vert >  \frac{1}{\vert \tau_{\rm in} \vert} \;\;\;,\;\;\;
\vert \tau \vert < \vert \tau_i \vert <  {\rm Min} \left[ \frac{1}{  \vert \vec{p}_i \vert } ,\, \frac{1}{  \vert {\vec k}_i - \vec{p}_i \vert } \right]  
\label{condition3}
\end{equation}

Performing the time integrals and employing the $\delta-$functions, we obtain
\begin{eqnarray}
\left\langle \prod_{i=1}^3  \frac{\delta \varphi_{\vec k_i} }{a}  \right\rangle_\tau \vert_{\rm sourced}
&\simeq& 
\frac{ 27 }{8 \pi^{9/2} } \frac{H^6}{\epsilon^{3/2} M_p^3} \delta^{(3)} \left( \vec{k}_1 + \vec{k}_2 + \vec{k}_3 \right) \nonumber\\
&&\!\!\!\!\!\!\!\!\!\!\!\!\!\!\!\!\times \,  \int d^3 p \, \ln {\rm Min} \left[ \frac{1}{\vert \vec{p} \vert \, \vert \tau \vert} ,\, \frac{1}{\vert \vec{k}_1 - \vec{p} \vert \, \vert \tau \vert} \right]
 \, \ln {\rm Min} \left[  \frac{1}{\vert \vec{k}_1 - \vec{p} \vert \, \vert \tau \vert}  ,\,  \frac{1}{\vert \vec{k}_3 + \vec{p} \vert \, \vert \tau \vert} \right] \,
 \ln {\rm Min} \left[  \frac{1}{\vert \vec{k}_3 + \vec{p} \vert \, \vert \tau \vert}  ,\,  \frac{1}{\vert  \vec{p} \vert \, \vert \tau \vert} \right] 
\nonumber\\
&&\times
\frac{\sum_{\lambda_1} \epsilon_k^{(\lambda_1)*} \left( \vec{p} \right) \,  \epsilon_i^{(\lambda_1)} \left( \vec{p} \right) \,  \sum_{\lambda_2} \epsilon_i^{(\lambda_2)*} \left( \vec{p} - \vec{k}_1  \right) \,  \epsilon_j^{(\lambda_2)} \left( \vec{p} -  \vec{k}_1 \right)  \,  \sum_{\lambda_3} \epsilon_j^{(\lambda_3)*} \left( \vec{p} + \vec{k}_3  \right) \,  \epsilon_k^{(\lambda_3)} \left( \vec{p} +  \vec{k}_3 \right) }{ \vert \vec{p} \vert^3 \, \vert \vec{p} - \vec{k}_1 \vert^3 \,  \, \vert \vec{p} + \vec{k}_3 \vert^3 } 
\nonumber\\
\label{res3-par}
\end{eqnarray}
where, due to (\ref{condition3}), the integration region is delimited by
\begin{equation}
\vert \vec{p} \vert ,\; \vert \vec{p} - \vec{k}_1 \vert ,\; \vert \vec{p} + \vec{k}_3 \vert > \frac{1}{\vert \tau_{\rm in} \vert}
\label{condition4}
\end{equation}

The momentum  integral in (\ref{res3-par}) has most of its support at the three logarithmic poles; each pole occurs when one of the three quantities in (\ref{condition4}) reaches its minimal value $\frac{1}{\vert \tau_{\rm in} \vert}$. Formally,
\begin{equation}
\left\langle \prod_{i=1}^3  \frac{\delta \varphi_{\vec k_i} }{a}  \right\rangle_\tau \vert_{\rm sourced}
\simeq {\cal C}_{\vert \vec{p} \vert \simeq \frac{1}{\vert \tau_{\rm in} \vert}} \left[ \vec{k}_1 ,\, \vec{k}_2 ,\, \vec{k}_3 \right] +  {\cal C}_{\vert \vec{p} - \vec{k}_1  \vert \simeq \frac{1}{\vert \tau_{\rm in} \vert}} \left[ \vec{k}_1 ,\, \vec{k}_2 ,\, \vec{k}_3 \right] + {\cal C}_{\vert \vec{p} + \vec{k}_3  \vert \simeq \frac{1}{\vert \tau_{\rm in} \vert}} \left[ \vec{k}_1 ,\, \vec{k}_2 ,\, \vec{k}_3 \right] 
\end{equation}
where ${\cal C}$ refers to the contribution to the integral in  (\ref{res3-par}) from the region close to the pole indicated by the suffix.

To evaluate the contribution from the second region, we redefine  the integration variable as $\vec{p} \rightarrow \vec{p} + \vec{k}_1$; we then see that this contribution is formally equal to the contribution from the first region, provided the external momenta in the first region are changed as $\vec{k}_1 \rightarrow \vec{k}_2$, $\vec{k}_2 \rightarrow \vec{k}_3$, and $\vec{k}_3 \rightarrow \vec{k}_1$.
Analogously, to  evaluate the contribution from the third region, we redefine  the integration variable as $\vec{p} \rightarrow \vec{p} - \vec{k}_3$; we then see that this contribution is formally equal to the contribution from the first region, provided the external momenta in the first region are changed as $\vec{k}_1 \rightarrow \vec{k}_3$, $\vec{k}_2 \rightarrow \vec{k}_1$, and $\vec{k}_3 \rightarrow \vec{k}_2$.
We therefore have
\begin{equation}
\left\langle \prod_{i=1}^3  \frac{\delta \varphi_{\vec k_i} }{a}  \right\rangle_\tau \vert_{\rm sourced}
\simeq {\cal C}_{\vert \vec{p} \vert \simeq \frac{1}{\vert \tau_{\rm in} \vert}} \left[ \vec{k}_1 ,\, \vec{k}_2 ,\, \vec{k}_3 \right] + {\rm permutations}
\end{equation}

To evaluate this contribution, we use the identity
\begin{equation}
\sum_\lambda \epsilon_i^{(\lambda)*} \left( \vec{p} \right) \,  \epsilon_j^{(\lambda)} \left( \vec{p} \right) = 
\delta_{ij} - {\hat p}_i \, {\hat p}_j
\end{equation}
and we obtain
\begin{eqnarray}
\left\langle \prod_{i=1}^3  \frac{\delta \varphi_{\vec k_i} }{a}  \right\rangle_\tau \vert_{\rm sourced}
& \simeq &
\frac{ 27 }{8 \pi^{9/2} } \frac{H^6}{\epsilon^{3/2} M_p^3} \delta^{(3)} \left( \vec{k}_1 + \vec{k}_2 + \vec{k}_3 \right)  \nonumber\\
&&
\times \, 
\Bigg\{
\ln \frac{1}{k_1 \, \vert \tau \vert} \, \ln \frac{1}{k_3 \, \vert \tau \vert} \, 
\ln {\rm Min} \left[  \frac{1}{k_1 \, \vert \tau \vert} ,\,  \frac{1}{k_3 \, \vert \tau \vert}  \right] 
 \frac{8 \pi}{3} \left[ 1 + {\rm cos }^2 \left( \vec{k}_1 ,\, \vec{k}_3 \right) \right] \, \frac{1}{k_1^3 k_3^3} \, {\rm ln } \, {\rm Min} \left[ k_1 \, \vert \tau_{\rm in} \vert ,\, k_3 \, \vert \tau_{\rm in} \vert \right] \nonumber\\
&&
 \quad\quad\quad\quad\quad\quad\quad\quad
  + {\rm permutations}
\Bigg\} 
\label{full-ki}
\end{eqnarray}

Assuming that the external momenta are not too hierarchical (see below), we disregard the difference among them in the argument of the logarithms. The expression for the correlator, evaluated at the end of inflation,  then simplifies to
\begin{eqnarray}
\left\langle \prod_{i=1}^3  \frac{\delta \varphi_{\vec k_i} }{a}  \right\rangle_{\tau_{\rm end}} \vert_{\rm sourced}
& \simeq &
144 \, \sqrt{\frac{2}{\pi}} \, {\cal P}^{3/2}  \, H^3   \,  \delta^{(3)} \left( \vec{k}_1 + \vec{k}_2 + \vec{k}_3 \right) N_{\rm CMB}^3 \left( N_{\rm tot} - N_{\rm CMB} \right) \nonumber\\
&&\quad\quad\quad\quad \times
\left[ \frac{1+\cos^2 \left( \vec{k}_1 ,\, \vec{k}_2 \right)}{ k_1^3 \, k_2^3} +  \frac{1+\cos^2 \left( \vec{k}_1 ,\, \vec{k}_3 \right)}{ k_1^3 \, k_3^3}  +  \frac{1+\cos^2 \left( \vec{k}_2 ,\, \vec{k}_3 \right)}{ k_2^3 \, k_3^3} 
\right] \nonumber\\
\end{eqnarray}
We can now evaluate the 3-point function of the curvature perturbation using the relation $\zeta = -\frac{H}{\dot{\varphi}_0}\delta\varphi$ and introduce a momentum-dependent nonlinearity parameter by comparison with (\ref{local_bi_first}).  We find:
\begin{eqnarray}
f_{NL}(k_i) &\simeq& f_{NL}^{\rm equiv. \;  local} \times \frac{3}{4} \,  \frac{ \frac{1+\cos^2 \left( \vec{k}_1 ,\, \vec{k}_2 \right)}{ k_1^3 \, k_2^3} +  \frac{1+\cos^2 \left( \vec{k}_1 ,\, \vec{k}_3 \right)}{ k_1^3 \, k_3^3}  +  \frac{1+\cos^2 \left( \vec{k}_2 ,\, \vec{k}_3 \right)}{ k_2^3 \, k_3^3} }{\frac{1}{k_1^3 \, k_2^3} + \frac{1}{k_1^3 \, k_3^3} + \frac{1}{k_2^3 \, k_3^3} } \nonumber\\
 f_{NL}^{\rm equiv. \;  local}  &\simeq&  1280 \, \frac{\mathcal{P}^3}{P_\zeta(k)^2} N_{\rm CMB}^3 \left( N_{\rm tot} - N_{\rm CMB} \right) \label{fNLfinalderived}
\end{eqnarray}
If the power spectrum is dominated by the vacuum fluctuations (which we assume for this work) then we have $P_\zeta(k) \approx \mathcal{P}$ and $f_{NL}^{\rm equiv. \;  local}  \simeq  1280 \, \mathcal{P} N_{\rm CMB}^3 \left( N_{\rm tot} - N_{\rm CMB} \right)$.

We conclude by noting that a more precise shape than (\ref{fNLfinalderived}) can be readily obtained from (\ref{full-ki}) also for hierarchical momenta; for instance, in the limit $k_1 \ll k_2 \simeq k_3$, the third term in the numerator of $f_{NL}(k_i) $ becomes irrelevant, and the $N_{\rm CMB}$ factors entering in $ f_{NL}^{\rm equiv. \;  local} $ should read $N_{{\rm CMB},1} \, N_{{\rm CMB},2}^2 \, \left( N_{\rm tot} - N_{{\rm CMB},1} \right)$, where $N_{{\rm CMB},i}$ refers to the horizon-exit of the mode with momentum $k_i$.

\section{Tensor modes}
\label{sec:tensor}

Production of gauge field fluctuations during inflation and its effect on the curvature perturbation have been discussed in the previous Sections. The produced gauge quanta, however, affect not only the scalar but also tensor perturbations (gravity waves). Metric perturbations couple to each content of the universe and are inevitably sourced by the produced gauge field fluctuations. To see the effect, we consider the transverse and traceless components of the spatial metric perturbations: $g_{ij} = a^2 \left( \delta_{ij} + h_{ij} \right)$, with $h_{ii} = 0$ and $\partial_i h_{ij} = 0$. As the matter content is scalar and vector, $h_{ij}$ is the only tensor perturbations in the model.

From the Einstein equations, one finds the same equations for $h_{ij}$ in terms of the physical $E_i$ and $B_i$ as in \cite{Barnaby:2011vw},
\begin{equation}
\frac{1}{2 a^2} \left( \partial_\tau^2 + 2 \frac{a'}{a} \partial_\tau - \triangle \right) h_{ij} = - \frac{1}{M_p^2} \left( E_i E_j + B_i B_j \right)^{TT} \label{eq_hij}
\end{equation}
where $TT$ denotes the transverse and traceless projection of the spatial components of the energy-momentum tensor of the gauge field.%
\footnote{This projection can be done by an operator ${\mathcal O}_{ij , lm}$, which can be expressed in the momentum space as ${\mathcal O}_{ij , lm} \big( \hat{k} \big) = P_{il} \big( \hat{k} \big) P_{jm} \big( \hat{k} \big) - \frac{1}{2} P_{ij} \big( \hat{k} \big) P_{lm} \big( \hat{k} \big)$, where $P_{ij} \big( \hat{k} \big) = \delta_{ij} - \hat{k}_i \hat{k}_j$.}
Since the gauge field has no expectation value, there is no coupling to the tensor perturbations at linearized level. Thus (\ref{eq_hij}) is in fact up to second order, and the right-hand side should in principle contain the source terms from squares of the first-order inflaton and metric  perturbations. As for $\delta_2 \varphi$, such source terms are uncorrelated and subdominant to those from the gauge field.

Tensor modes (or GW) are transverse and traceless part of $\delta g_{ij}$ and have two physical degrees of freedom, the left-handed ($L$) and right-handed ($R$). It is convenient to decompose tensor perturbations as
\begin{equation}
h_{ij} \left( \tau , \vec{x} \right) = \int \frac{d^3k}{\left( 2 \pi \right)^{3/2}} \, {\rm e}^{i \vec{k} \cdot \vec{x}} \sum_{s = L , R} \Pi_{ij , s} \big( \vec{k} \big) \hat{h}_s \big( \vec{k} \big)
\end{equation}
where $\hat{h}_s \big( \vec{k} \big) = h_s \left( k \right) a_{s}({\vec{k}}) + h_s^* \left( k \right) a_{s}^{\dagger}(-\vec{k})$, and the helicity projectors are $\Pi_{ij , L/R} \big( \vec{k} \big) = e_i^{\left( \mp \right)} \big( \vec{k} \big) \, e_j^{\left( \mp \right)} \big( \vec{k} \big)$, which clearly have the properties $\Pi_{ii , s} \big( \vec{k} \big) = \hat{k}_i \, \Pi_{ij , s} \big( \vec{k} \big) = 0$. Note that $h_s \left( k \right)$ depends only on the magnitude of $\vec{k}$. Since the mechanism of GW production is analogous to that of the curvature perturbations, presented in detail in the previous Sections, we merely show the result of our computation here. Define GW power spectrum $P_{L/R}$ in the usual way,
\begin{equation}
\left< \hat{h}_s \big( \vec{k} \big) \, \hat{h}_s \left( \vec{k}' \right) \right> \equiv P_s \left( k \right) \, \frac{2 \pi^2}{k^3} \, \delta^{(3)} \big( \vec{k} + \vec{k}' \big)
\end{equation}
where $s = L/R$. As for the scalar perturbations, the GW modes produced by the gauge quanta are uncorrelated with those from vacuum fluctuations, and so the two contributions simply add up in the power spectrum.  The two helicity states are produced in the same amount, and their sum gives
\begin{equation}
P_{\mathrm{GW}} \left( k \right) \cong \frac{2H^2}{\pi^2 M_p^2} \left[ 1 + \frac{6 H^2}{\pi^2 M_p^2} \ln^2 \frac{a \left( \tau \right) H}{k} \, \ln \frac{k}{a \left( \tau_{\rm{in}} \right) H} \right]
\end{equation}
where the first term in the square brackets is the contribution from the vacuum and the latter from the source.  Evaluating this expression at the end of inflation gives
\begin{equation}
  \left. P_{\mathrm{GW}} \right|_{\mathrm{end}\hspace{1mm}\mathrm{inflation}} 
  = \frac{2H^2}{\pi^2 M_p^2} \left[ 1 + \frac{6 H^2}{\pi^2 M_p^2} N_{\mathrm{CMB}}^2 \left(N_{\mathrm{tot}}-N_{\mathrm{CMB}}\right) \right]
\end{equation}
As expected, the standard vacuum part dominates in the regime we are interested in:  when the vacuum contribution to $P_\zeta$ is dominant over that from the source, the same is also true for the GW power spectrum; this is due to the fact that the tensor modes are only produced gravitationally, while the dominant source of the scalar modes is the direct inflaton-gauge field coupling; (this coupling is mathematically enhanced with respect to the gravitational one by  $1/\epsilon$). In this regime, the tensor-to-scalar ratio $r$ reproduces the standard result:\footnote{We define the tensor-to-scalar ratio in the usual way, by normalizing the power in GW to that in curvature perturbations: $r \equiv P_{GW}/P_\zeta = \left( P_L + P_R \right)/P_\zeta$.}
\begin{equation}
 r \approx 16 \epsilon
\end{equation}

\section{Validity of Perturbation Theory}
\label{sec:weyl}

It is important to verify that cosmological fluctuations remain small in our model, to justify a perturbative analysis.  In this Section we consider several different diagnostics, arguing that perturbation theory is well under control, and that the second order results we have computed are subdominant with respect to the standard first order results (apart of course for what concerns the non-gaussian nature of $\zeta$, which is not present in the linearized theory).

As discussed previously, the produced gauge field fluctuations backreact on the classical background, by contributing to the total energy density in the Friedmann equation and by introducing dissipation into the equation for the homogeneous inflaton.  A first consistency check is to ensure that these effects are negligible, which we have already shown in Section \ref{sec:EB}.

Another important diagnostic is the variance of curvature fluctuations, $\langle \zeta^2 \rangle^{1/2}$ (in general, we need to compute the variance of a perturbation as a measure of the value of that perturbation in real space). Using our previous result for the 2-point correlation function in momentum space, it is straightforward to compute the total variance of curvature fluctuations
\begin{eqnarray}
 \langle \zeta^2(\tau,{\vec x}) \rangle &=&  \left. \langle \zeta^2(\tau,{\vec x}) \rangle \right|_{\mathrm{vac}} 
   + \,\,\,\left. \langle \zeta^2(\tau,{\vec x}) \rangle \right|_{\mathrm{sourced}}  \, , \nonumber \\
  &\cong&  \mathcal{P} \int \frac{dk}{k}  \,\,\,\,\,\,\,\,\,\,\,\,
           +\,\,\,\,192\mathcal{P}^2 \int \frac{dk}{k} \ln^2\left(\frac{a(\tau) H}{k}\right)\ln\left(\frac{k}{a_{\mathrm{in}}H}\right) \, , \nonumber \\
  &\cong& \mathcal{P} \ln\left(\frac{a(\tau)}{a_{\mathrm{in}}}\right)\left[ 1 + 16 \mathcal{P} \ln^3\left(\frac{a(\tau)}{a_{\mathrm{in}}}\right)  \right] \, .
\label{zetavar}
\end{eqnarray}
To compute (\ref{zetavar}) we have integrated from $k_{\mathrm{min}} = a_{\mathrm{in}} H$ to $k_{\mathrm{max}} = a(\tau) H$, since we should count the phase space of super-horizon modes from the onset of inflation to the time $\tau$.  Equation (\ref{zetavar}) shows that $\langle \zeta^2 \rangle^{1/2} \ll 1$ generically and, moreover, contributions from sourcing effects are controlled by the small parameter $\mathcal{P} \sim 10^{-9}$.

Since the spectrum of sourced curvature fluctuations exhibits a logarithmic growth, one might worry about estimators of inhomogeneity that contain derivatives of the fluctuations, for example the energy density or the Weyl curvature.  We consider such diagnostic in the next two Subsections, which are devoted to the contributions from tensor and scalar modes, respectively. In the final Subsection we conclude that perturbation theory is well under control.

\subsection{Tensor Modes}

We first consider the contribution to the energy density and Weyl tensor from tensor perturbations, since they are technically simpler to evaluate.  

The energy density in gravitational waves contains a contribution due to the usual vacuum fluctuations, and an uncorrelated contribution from sourcing effects.  The former is given by
\begin{eqnarray}
 \left. \delta\rho_{\mathrm{GW}} \right|_{\mathrm{vac}} &=& \frac{M_p^2}{16\pi^2 a^2} \sum_\lambda \int dk k^2 \left[ |h_\lambda'(k)|^2 + k^2 |h_\lambda(k)|^2  \right]_{\rm vac}         \nonumber \\
 &\approx& \frac{H^2}{4\pi^2}\frac{1}{a^2} \int \frac{dk}{k} \left[ k^2 + 2 k^4 \tau^2 \right] 
 \approx  \frac{H^4}{4 \pi^2} \label{GWvacE}
\label{GW-rho-vac}
\end{eqnarray}
The sourced contribution is given by
\begin{eqnarray}
 \left. \delta\rho_{\mathrm{GW}} \right|_{\mathrm{sourced}} &\approx& \frac{H^2}{4\pi^2}\frac{1}{a^2} \int \frac{dk}{k} 
  \left[  \frac{6H^2}{\pi^2 M_p^2} \ln\left(\frac{k}{a_{\mathrm{in}}H}\right)\left( \frac{1}{\tau^2} + k^2 \ln^2\left(\frac{a(\tau) H}{k}\right) \right)   \right] \nonumber \\
  &\approx& \frac{3H^6}{4 \pi^4M_p^2} \ln^2\left(\frac{a(\tau)}{a_{\mathrm{in}}}\right) \label{GWsourceE} 
\label{GW-rho-source}
\end{eqnarray}
In both (\ref{GWvacE}) and (\ref{GWsourceE}) we have integrated from $k_{\mathrm{min}} = a_{\mathrm{in}} H$ to $k_{\mathrm{max}} = a(\tau) H$, as discussed above.  We see that the sourced contribution is subdominant to the vacuum one, being suppressed by the small number $H^2 / M_p^2 \sim 10^{-10}$.  The constraint $\delta\rho_{\mathrm{GW}} \ll 3 H^2 M_p^2$ is easily satisfied.

It is interesting to compare the time dependence in the final results (\ref{GWvacE}) and (\ref{GWsourceE}) with what would be obtained by inspecting the spectral densities.  In position space, the sourced contribution exhibits a logarithmic growth, relative to the vacuum contribution (we note that, however, it remains always very subdominant).  For the spectral densities, on the other hand, we see that the results differ by a power-law dependence on $\tau$ (this does not imply that either result is growing with time; indeed, neither term grows,  due to the time dependence of the scale factor outside the integral). This difference in behavior is not seen in the integrated result,  when the phase space of relevant modes is accounted for.

The Weyl tensor is a gauge invariant object which vanishes for an FRW universe and thus may be considered as a measure of inhomogeneities.  A good estimator for the size of typical entries in the Weyl tensor (from gravity wave perturbations) is the quantity
\begin{eqnarray}
  C^T(\tau,{\vec x}) \equiv  h'' + \Lap h  \, .
\end{eqnarray}
Refs. \cite{Bonvin:2011dt} used as a dimensionless diagnostic of the validity of perturbation theory 
the ratio between the entries of this tensor and  those of the Ricci tensor, whose background components are of the order $\sH^2$.  Proceeding as above, the contribution from vacuum fluctuations is
\begin{eqnarray}
 \sH^{-4} \left. \langle (C^T(t,{\vec x}))^2 \rangle \right|_{\mathrm{vac}} &\approx& 
 \frac{\tau^4}{2\pi^2} \int\frac{dk}{k} \left[  \frac{8H^2k^6\tau^2}{M_p^2}  \right] 
 \approx \frac{2}{3\pi^2}\frac{H^2}{M_p^2} 
\label{tensor-Weyl-vac}
\end{eqnarray}
For the sourced contribution we instead have
\begin{eqnarray}
 \sH^{-4} \left. \langle (C^T(t,{\vec x}))^2 \rangle \right|_{\mathrm{sourced}} &\approx&
 \frac{\tau^4}{2\pi^2} \int\frac{dk}{k} \left[ \frac{12 H^4}{\pi^2 M_p^4} \ln\left(\frac{k}{a_{\mathrm{in}}H}\right) \left(  \frac{1}{\tau^2} - k^2 \ln\left(\frac{a(\tau) H}{k}\right)  \right)^2  \right] \nonumber \\
  &\approx& \frac{3 H^4}{\pi^4 M_p^4}\ln^2\left(\frac{a(\tau)}{a_{\mathrm{in}}}\right) 
\label{tensor-weyl}
\end{eqnarray}
Again, we see that the contribution from sourcing effects exhibits a logarithmic growth, but remains safely subdominant to the vacuum part.  As for the energy density, the  power-law time dependence in the spectral density is compensated by phase space factors.

\subsection{Scalar Modes}

We now consider contributions to the energy density and Weyl tensor from scalar fluctuations.  The energy in inflaton fluctuations, $\delta \rho_\varphi =  \frac{1}{2a^2} \langle (\delta\varphi')^2 + (\vec{\nabla}\delta\varphi)^2 + a^2 V_{,\varphi\varphi} (\delta\varphi)^2  \rangle$, is given by 
\begin{eqnarray}
  \delta\rho_{\varphi} &=& \left.\delta\rho_\varphi\right|_{\mathrm{vac}}  +   \left.\delta\rho_\varphi\right|_{\mathrm{sourced}}        \, , \nonumber \\
  &\cong& \frac{H^4}{8\pi^2} \left\{ \,\, \left[ 1 + 3\eta \ln\left(\frac{a(\tau)}{a_{\mathrm{in}}}\right) \right]\,\, + \,\,96 \mathcal{P} \ln^2\left(\frac{a(\tau)}{a_{\mathrm{in}}}\right) \left[ 1 + \frac{\eta}{2}\ln^2\left(\frac{a(\tau)}{a_{\mathrm{in}}}\right) \right]\,\, \right\} \, ,
\end{eqnarray}
where the first term is the usual contribution from the vacuum, the second term arises due to the feed-back of produced gauge fluctuations into the inflaton perturbations, and $\eta \approx V_{,\varphi\varphi} / (3 H^2)$ is the standard slow roll parameter. We see that the condition $\delta \rho_{\varphi} \ll 3 H^2 M_p^2$ is easily satisfied, and that the sourced contribution is subdominant.
 
A good estimator for the size of the contribution of the  scalar fluctuations  to the Weyl tensor is the quantity
\begin{equation}
\label{CSdef}
  C^S \equiv \Lap \left(\Phi + \Psi\right)
\end{equation}
where $\Phi$ and $\Psi$ are the gauge invariant Bardeen potentials.  This estimator is discussed in some detail in Appendix A.  Again, we compare this to $\sH^2$ to have a dimensionless diagnostic of the validity of perturbation theory.  The contribution to our estimator (\ref{CSdef}) from the usual vacuum fluctuations arises at linear order in perturbation theory 
\begin{equation}
\label{CSvac}
 \left. C^S(\tau,{\vec x}) \right|_{\mathrm{vac}} \approx \epsilon (\zeta_1'' - \Lap \zeta_1)
\end{equation}
It is straightforward to compute the variance
\begin{eqnarray}
  \sH^{-4} \left. \langle \left( C^S(\tau,{\vec x}) \right)^2 \rangle \right|_{\mathrm{vac}} &=& \epsilon^2 \sH^{-4} \langle (\zeta_1'' - \Lap \zeta_1)^2 \rangle \, , \nonumber \\
  &\cong& \epsilon^2\tau^4 \int\frac{d^3k}{(2\pi)^3} \left| u_k'' + k^2 u_k  \right|^2 \, , \nonumber \\ 
  &\cong& \epsilon^2 \mathcal{P} \, . \label{WeylVac}
\end{eqnarray}
where in the second line we have introduced the notation $u_k(\tau) \equiv - i \pi^{3/2}\mathcal{P}^{1/2} (-\tau)^{3/2} H_{3/2}^{(1)}(-k\tau)$ for the mode
functions.  Before moving on, it is worth commenting on the constancy of the result (\ref{WeylVac}).  
Even if the Fourier transform of $\Lap \zeta / \sH^2$ is decreasing in time in the super-horizon regime,  the momentum integral that gives the variance (\ref{WeylVac}) is 
dominated by the UV cut-off $k=aH$; this phase space factor compensates for the factor $\sH^{-2} = (aH )^{-2}$.  (Note that this same phase space 
compensation was seen also in Section \ref{sec:EB} when we computed the energy density in electric and magnetic fields, showing that $\rho_E$ is 
``only'' logarithmically suppressed with respect to $\rho_B$, even though the electric field Fourier modes decay as $a^{-1}$ outside the horizon.)  Despite not decreasing in time, the integrated result is  $\ll 1$.

At second order in perturbation theory there arises also a contribution to the Weyl tensor due to the feed-back of the produced gauge fluctuations on the 
inflaton and scalar metric perturbations.  In Appendix A we show that this is of the form
\begin{equation}
\label{CSsource}
 \left. C^S(\tau,{\vec x}) \right|_{\mathrm{sourced}} = \epsilon (\zeta_2'' - \Lap \zeta_2) + \Upsilon \, .
\end{equation}
The source term $\Upsilon$ is quadratic in gauge field fluctuations and the full expression (given in Appendix A) is quite cumbersome.  For our purposes, 
it suffices to consider a representative contribution to the source term
\begin{equation}
 \Upsilon \sim \frac{a^2}{M_p^2} \left[ \vec{E}^2 + \vec{B}^2 \right] + \cdots 
\end{equation}
During inflation, such quantities are tiny as compared to $\sH^2$ by virtue of the backreaction constraint $\rho_{E} + \rho_{B} \ll 3 H^2 M_p^2$.  Moreover, 
notice that the source term $\Upsilon$ will quickly decay as $a^{-2}$ after inflation has ended (at which point the production of gauge fluctuations comes 
to a halt and we recover the usual behavior $\rho_{E} \sim \rho_{B} \sim a^{-4}$).  By contrast, we expect that $\zeta_2$ becomes frozen on large scales 
after inflation.  Therefore the source term $\Upsilon$ in (\ref{CSsource}) should be irrelevant for any late-time observable.

Let us now compute the variance of the first term in (\ref{CSsource}).  Following very closely our previous computations we find
\begin{equation}
 \sH^{-4} \left. \langle \left( C^S(\tau,{\vec x})  \right)^2 \rangle \right|_{\mathrm{sourced}} 
    \cong  96\epsilon^2 \mathcal{P}^2 \ln^2\left(\frac{a(\tau)}{a_{\mathrm{in}}}\right)  + \cdots
\end{equation}
This exhibits a logarithmic growth during inflation but it always remains much smaller than the vacuum contribution.

\subsection{Summary}

To summarize, in this Section we have computed the contribution to various physical quantities from both the usual vacuum fluctuations and also from the feed-back of produced gauge field fluctuations on the inflaton and scalar metric perturbations.  We have found that the ``new'' contributions from sourcing effects generically exhibit logarithmic growth during inflation and remain safely sub-dominant with respect to the standard result.  This statement applies to the curvature perturbation, the energy density in fluctuations, and the Weyl tensor. We have computed our diagnostics during inflation.  In principle, these results should be evolved through reheating and into the radiation phase, to ensure that the universe remains homogeneous and isotropic after the end of inflation.  We do not expect any problem, since all the physical quantities  at the end of inflation are dominated by the usual vacuum contribution (which does not lead to any unacceptable instability), since the energy density in the gauge field is subdominant, and since we assume a standard $\sqrt{-g} F^2$ term from the end of inflation onwards. 

Ref.~\cite{Bonvin:2011dt} argued that the gauge field production in these frameworks leads to  an unacceptable breakdown of homogeneity and isotropy soon after inflation, after the perturbations are matched to the post-inflationary epoch. \footnote{These comments refer to version $1$ of  \cite{Bonvin:2011dt} as it appeared on the public archive.  See our comments in the Note Added at the end of this manuscript.}  We believe that the core of their claim is in the time dependent behavior of the spectral density of the sourced part of the Weyl tensor during inflation; they pointed out this feature in the scalar sector, but we performed computations also in the tensor sector because the identical feature appears also there, and the computations are simpler to present.  We note that indeed there is a different time dependence in the spectral densities of eq. (\ref{GW-rho-vac}) vs.  (\ref{GW-rho-source})  and of eq. (\ref{tensor-Weyl-vac}) vs.  (\ref{tensor-weyl}). We note, however, that this does not imply that the sourced term dominates over the vacuum one in physical integrated  quantities like the energy density and the Weyl tensor (if the dominance does not take place before the end of inflation, it will not take place afterwards). All the sourced quantities that we have studied remain small both in real and momentum space, and do not grow as power law of the scale factor.

\section{Conclusion}
\label{sec:conclusions}

In this paper, we have considered a simple model where the scalar inflaton is coupled to some $U(1)$ gauge field in a way that would be typical for moduli or dilaton-like fields.  We have seen that the time dependence of the condensate during inflation leads to a production of large scale gauge field fluctuations, analogous to the usual mechanism that amplifies the quantum vacuum fluctuations of the inflaton or gravity wave perturbations.  Focusing on the case where the spectrum of produced ``magnetic'' fields is scale invariant, we have shown that (nearly) local nongaussianity is very naturally generated at the level $f_{NL} \gsim \mathcal{O}(10)$, which should be probed in the near future.  Although we have neglected slow roll corrections to the running of the spectrum and bispectrum, we can still see a logarithmic running of the effective $f_{NL}$ parameter with scale.  This arises since the ``sourced'' contribution to the curvature perturbation experiences a super-horizon evolution during inflation, due to the presence of large scale iso-curvature perturbations.  Logarithmic running of this type may be of observational interest; see \cite{LoVerde:2007ri,Kumar:2009ge,Shandera:2010ei} for example.  Moreover, since the nongaussian part of $\zeta$ is uncorrelated with the gaussian part, we have a non-hierarchical scaling which can lead to interesting signatures in probes that are sensitive to the global structure of the PDF \cite{Barnaby:2011pe}.

A novel feature of our result is the dependence of $f_{NL}$ on the $N_{\mathrm{tot}} - N_{\mathrm{CMB}}$ which measures the number of ``extra'' e-foldings of inflation, beyond the minimal $N_{\mathrm{CMB}}\sim 60$ e-foldings between the end of inflation and horizon exit for CMB scales.  Ordinarily, one would expect that such ``extra'' e-foldings are completely unobservable, since scalar modes which leave the horizon prior to $N_{\mathrm{CMB}}$ should just be absorbed into a renormalization of the homogeneous background.  However, in our model the total duration of the quasi de Sitter phase impacts the energy density of produced gauge field \cite{Demozzi:2009fu}.  Indeed, if $N_{\mathrm{tot}}$ is too large then the backreaction of produced gauge fields will become appreciable and spoil inflation.  In a restricted sense, we see that in this model the ``extra'' e-foldings of inflation can influence physical observables.

We have seen that the correlation functions of scalar and tensor cosmological perturbations exhibit a logarithmic time dependence which is related to the growing phase space of produced gauge fluctuations.  These logarithms should not be confused with the IR logs that have been discussed extensively in the literature in association with loop effects during inflation \cite{Weinberg:2005vy,Seery:2007we,Senatore:2009cf,Burgess:2009bs} (see also \cite{Kumar:2009ge} for a related discussion).  In the case at hand the interpretation of this logarithmic time dependence is straightforward.  The production of gauge field fluctuations in our model arises simply because the effective gauge coupling is time dependent.  This time dependence leads to a growth in the energy density of gauge fluctuations, which is drained from the scalar condensate.  The energy transfer provides a physical clock, and its logarithmic growth is a real physical effect that is not related to the de Sitter background or to the quantization of gravitational fluctuations. 

There are a number of interesting directions for future research.  It may be for example interesting to study effects that can arise if the vector field has a vacuum expectation value already at zeroth order in perturbation theory. It would also be interesting to consider different choices of coupling function, for example $n\not= 2,-2$.  In connection with this, it would be interesting to explore in more detail the nongaussian phenomenology of our model, in particular as regards higher moments, scale dependence, and LSS probes.  It may also be possible to obtain interesting signals for gravitational waves at interferometers. Finally, it would be worth to investigate whether this mechanism for the amplification of the gauge ``magnetic'' modes can be consistently modified so to  avoid the strong coupling problem of \cite{Demozzi:2009fu}, and therefore be used as a mechanism for magnetogenesis.

\vspace{1cm}

%
%
%
%
{\bf Note added}: The stability analysis performed in Section \ref{sec:weyl} disagrees with the results of version 1 of \cite{Bonvin:2011dt}, as it appeared on the public archive.  After the present manuscript was posted on the archive, ref. \cite{Bonvin:2011dt} was replaced by a second version, which agrees with our conclusions that perturbations theory is under control.

\section*{Acknowledgments}

We thank N.~Bartolo, J.~Cline, T. Jones, A.~Lewis, K.~Malik, S.~Matarrese, K.~A.~Olive, L. Rudnick, D.~Seery, D.~Tran, M.~Voloshin, and M.~Wyman  for interesting correspondence and discussions. This work was supported in part by DOE grant DE-FG02-94ER-40823 at UMN. 

\appendix

\section{Cosmological Perturbation Theory}
\label{appA}

We work in the spatially flat gauge and expand the metric up to second order in perturbation theory as
\begin{eqnarray}
  g_{00} &=& -a^2 (1 + 2\phi_1 + 2\phi_2) \, , \\
  g_{0i} &=& a^2 \partial_i ( B_1 + B_2 ) \, ,\\
  g_{ij} &=& a^2 \delta_{ij} \, .
\label{deco-metric}
\end{eqnarray}
Similarly, the scalar and gauge fields are expanded as
\begin{eqnarray}
 \varphi(t,{\vec x}) &=& \varphi_0(t) + \delta_1 \varphi(t,{\vec x}) + \delta_2 \varphi(t,{\vec x}) \, , \\
  A_\mu(t,{\vec x}) &=& \left(0 , \delta_1 A_i(t,{\vec x}) + \delta_2 A_i(t,{\vec x}) \right) \, .
\end{eqnarray}

The equation of motion for the inflaton field is
\begin{equation}
\label{KGapp}
 \frac{1}{\sqrt{-g}} \partial_\mu\left( \sqrt{-g} g^{\mu\nu} \partial_\nu \varphi \right) - \frac{dV}{d\varphi} = \frac{1}{4} \frac{d I^2}{d\varphi} F^{\mu\nu}F_{\mu\nu} \, .
\end{equation}
We expand (\ref{KGapp}) up to second order, using the Einstein constraint equations to eliminate the metric fluctuations in order to close 
the system.  This procedure has already been detailed in \cite{Malik:2006ir}; here we generalize those results to include also the presence 
of the gauge field.  (The analogous procedure for an axial coupling is discussed in \cite{Barnaby:2011vw}.)

At linear order in perturbations, the Klein-Gordon equation (\ref{KGapp}) gives
\begin{equation}
\label{KG1step}
 \delta_1\varphi'' + 2\sH \delta_1\varphi' - \Lap \delta_1\varphi + a^2 V_{,\varphi\varphi}\delta_1\varphi - \varphi_0' \phi_1' + 2a^2V_{,\varphi} \phi_1 - \varphi_0' \Lap B_1 = 0 \, .
\end{equation}
From the $0-0$ and $0-i$ Einstein equations we can derive the results
\begin{eqnarray}
 \sH \Lap B_1 + \frac{1}{2M_p^2} \left[ 2 a^2 V\phi_1 +\varphi_0'\delta_1\varphi' + a^2 V_{,\varphi} \delta_1\varphi \right] &=& 0 \, , \label{B1}\\
 \sH \phi_1 - \frac{1}{2M_p^2} \varphi_0' \delta_1\varphi &=& 0 \, . \label{phi1}
\end{eqnarray}
Using these to eliminate $\phi_1$ and $B_1$ from (\ref{KG1step}) we obtain
\begin{equation}
\label{KG1}
 \left[ \frac{\partial^2}{\partial\tau^2} + 2\sH \frac{\partial}{\partial\tau} - \Lap + \left( a^2 V_{,\varphi\varphi} - 3\frac{(\varphi_0')^2}{M_p^2}  \right) \right] \delta_1 \varphi = 0 \, ,
\end{equation}
where we have made use of the background equations and work to leading order in slow roll parameters.

At second order in perturbation theory the procedure is nearly identical, save for the appearance of source terms, $S^{(i)}$, which are quadratic in first order fluctuations.  From the Klein-Gordon equation (\ref{KGapp}) we have
\begin{eqnarray}
  \delta_2\varphi'' + 2\sH \delta_2\varphi' - \Lap \delta_2\varphi + a^2 V_{,\varphi\varphi}\delta_2\varphi - \varphi_0' \phi_2' + 2a^2V_{,\varphi} \phi_2 - \varphi_0' \Lap B_2 
           = S^{(1)} && \label{KG2step} \\ 
 S^{(1)} \equiv \frac{ I I_{,\varphi}}{a^2}  \, 
  \left[ \delta_1A_i' \delta_1A_i' - \partial_i \delta_1A_j\partial_i \delta_1A_j +  \partial_i \delta_1A_j\partial_j \delta_1A_i  \right] \, 
  + \cdots \,\,\,\,\,\,\,\,\,\,\,\,\,\,\,\,\,\,\,\,\,\, && \label{S1}
\end{eqnarray}
where $\cdots$ denotes terms of the form $(\delta_1\varphi)^2$ which will not be important for our computation.  These have already been studied in \cite{Malik:2006ir} and they are known to give a negligible contribution to nongaussianity \cite{Seery:2008qj}.  The relevant constraint equations are
\begin{eqnarray}
\sH\Lap B_2 + \frac{1}{2M_p^2} \left[ 2 a^2 V\phi_2 +\varphi_0'\delta_2\varphi' + a^2 V_{,\varphi} \delta_2\varphi \right] &=& S^{(2)}  \, , \label{B2} \\
 \sH \phi_2 - \frac{1}{2M_p^2} \varphi_0' \delta_2\varphi &=& S^{(3)} \, , \label{phi2}
\end{eqnarray}
where the source terms are given explicitly by
\begin{eqnarray}
 S^{(2)} &\equiv& -\frac{I^2}{4a^2M_p^2} \left[ \delta_1A_i'\delta_1A_i' + \partial_i \delta_1 A_j (\partial_i\delta_1A_j - \partial_j\delta_1A_i)  \right] + \cdots \\ 
 S^{(3)} &\equiv& \frac{I^2}{2a^2M_p^2}\Lap^{-1} \left[ \partial_i \delta_1 A_j' (\partial_i\delta_1A_j - \partial_j\delta_1A_i) +  \delta_1A_i'\Lap\delta_1A_i     \right] + \cdots
\label{S2-S3}
\end{eqnarray}
Again, we have suppressed terms of the form $(\delta_1\varphi)^2$.  Combining the constraints (\ref{B2}) and (\ref{phi2}) with (\ref{KG2step}) we can eventually arrive at the master equation
\begin{eqnarray}
 && \left[ \frac{\partial^2}{\partial\tau^2} + 2\sH \frac{\partial}{\partial\tau} - \Lap + \left( a^2 V_{,\varphi\varphi} - 3\frac{(\varphi_0')^2}{M_p^2}  \right) \right] \delta_2 \varphi 
     = a^2 \frac{I_{,\varphi}}{I} \left(\vec{E}^2 - \vec{B}^2\right)  \nonumber \\
 && \,\,\,\,\,\,\,  - \frac{a^2 \varphi'}{2\sH M_p^2} \left[ \frac{\vec{E}^2 + \vec{B}^2}{2} 
              + \frac{1}{a^4} \Lap^{-1}\partial_\tau\left(a^4 \vec{\nabla}\cdot (\vec{E}\times\vec{B})\right)         \right] + \cdots
\label{masterapp} 
\end{eqnarray}
where the physical ``electric'' and ``magnetic'' fields are $E_i \equiv -\frac{\langle I \rangle}{a^2} \delta_1 A_i'$ and $B_i \equiv \frac{\langle I \rangle}{a^2} \epsilon_{ijk}\partial_j \delta_1 A_k$ respectively.  Equation (\ref{masterapp}) is the main result of this appendix.

Before finish, we briefly consider the Weyl tensor, which is discussed  in Section \ref{sec:weyl}.  The Weyl tensor from scalar perturbations has only an ``electric'' part which is given by
\begin{equation}
\label{Eij}
  E_{ij}= \frac{1}{2}\left[ \partial_i\partial_j - \frac{\delta_{ij}}{3}\Lap \right](\Phi + \Psi)
\end{equation}
where $\Phi$ and $\Psi$ are gauge invariant potentials, defined explicitly in \cite{Malik:2008im}.  A good estimator for the typical size
of entries in (\ref{Eij}) is the quantity
\begin{equation}
\label{CS}
  C^S \equiv \Lap (\Phi + \Psi) = \Lap \phi + \Lap B' \, ,
\end{equation}
where in the second equality we have restricted to the flat slicing.  

Equation (\ref{CS}) can be re-written in a suggestive manner.  Introducing the gauge invariant curvature perturbation 
$\zeta_n = -\frac{\sH}{\varphi_0'}\delta_n\varphi$ the constraint equations can be written in a convenient form
\begin{eqnarray}
  && \phi_1 = -\epsilon \zeta_1 \, , \hspace{5mm} \phi_2 = -\epsilon \zeta_2 + \sH^{-1} S^{(3)} \, , \\
  && \Lap B_1 = \epsilon \zeta_1' \, , \hspace{5mm} \Lap B_2 = \epsilon \zeta_2' + \sH^{-1} S^{(2)} - (3-\epsilon) S^{(3)} \, ,
\end{eqnarray}
at linear and quadratic order.  It is now straightforward to expand (\ref{CS}) in perturbation theory.  Working to leading order in slow roll parameters we find
\begin{eqnarray}
 C^S_1 &\cong& \epsilon (\zeta_1'' - \Lap \zeta_1) \, , \label{CS1} \\
  C^S_2 &\cong& \epsilon (\zeta_2'' - \Lap \zeta_2)  -(3-\epsilon) \partial_\tau S^{(3)} + \sH^{-1} \Lap S^{(3)} + \sH^{-1} \partial_\tau S^{(2)} - (1-\epsilon) S^{(2)} 
  \label{CS2} \, .
\end{eqnarray}

\section{Exact definition of $\zeta$}
\label{app:zeta}

In this Appendix we compute the exact expression for the gauge invariant curvature perturbation. We denote the exact expression by  $\zeta_{\rm exact}$. We instead denote by $\zeta$ the combination 
 $\zeta = - \frac{H}{\dot{\varphi}_0} \, \delta \varphi$, as we do everywhere in this paper. In this Appendix we show that the difference between $\zeta$ and  $\zeta_{\rm exact}$ is completely negligible for all our purposes. We decompose 
\begin{eqnarray}
\zeta_{\rm exact} &=& \zeta_{{\rm exact},1} +  \zeta_{{\rm exact},2} \nonumber\\
&=& \left(  \zeta_{{\rm exact},1}  + \zeta_{{\rm exact},2} \vert_{{\rm sourced \; by \; } \delta_1 \varphi \; {\rm and } \;  \delta_1 g_{\mu \nu}} \right)
+   \zeta_{{\rm exact},2}  \vert_{{\rm sourced \; by \; } \delta_1 A_\mu}  
\label{zeta-formal}
\end{eqnarray}
where the number in the suffix denotes the order in perturbation theory, and where, in total analogy with  what we did in (\ref{df-formal}), we have separated the part of $ \zeta_{{\rm exact},2}$ sourced by the first order vacuum fluctuations of the inflaton and the metric from the part sourced by the vector field
(we remind that $\delta_1 A_\mu$ coincides with the quantity denoted by $A_\mu$ in the main text, since the gauge field has no vacuum expectation value). The two terms in the round parenthesis are uncorrelated with the last term. These two terms coincide with those computed without gauge field (more precisely, the gauge field affects them only due to the backreaction on the background dynamics, which we impose to be subdominant). As in the standard case, for these terms we have at super-horizon scales
\begin{equation}
 \zeta_{{\rm exact},1}  + \zeta_{{\rm exact},2} \vert_{{\rm sourced \; by \; } \delta_1 \varphi \; {\rm and } \;  \delta_1 g_{\mu \nu}}  = - \frac{H}{\dot{\varphi}_0} \delta_1 \varphi +  \mathcal{O} \left[ \left( \delta_1\varphi\right)^2
,\, \left( \delta_1 g_{\mu \nu} \right)^2 ,\, \left( \delta_1\varphi \times  \delta_1 g_{\mu \nu} \right)
\right]
\end{equation}
The precise expression for the second term is given in  eq. (7.71) of \cite{Malik:2008im}. Since the standard single scalar field inflation results apply for these terms, we know that the quadratic term in 
this expression gives a negligible contribution to the power spectrum and leads to unobservable non-gaussianity. Therefore, we disregard it in this work. 

The formal expression for $ \zeta_{{\rm exact},2}  \vert_{{\rm sourced \; by \; } \delta_1 A_\mu}  $ can be immediately obtained from  eq. (7.71) of \cite{Malik:2008im}. This expression is written before fixing any gauge, and reads
\begin{equation}
\label{start}
\zeta_{{\rm exact},2} = - \psi_2 - \frac{\sH}{\rho_0'} \delta_2 \rho + \mathcal{O} \left[ \left( \delta_1\varphi \right)^2
,\, \left( \delta_1 g_{\mu \nu} \right)^2 ,\, \left( \delta_1\varphi \times  \delta_1 g_{\mu \nu} \right)
\right]
\end{equation}
where $\psi_2$ is a second order perturbation entering in the spatial part of the metric,
$\delta_2 g_{ij,{\rm scalar}} = a^2 \left[ - 2 \psi_2 \delta_{ij} + 2 E_{2,ij} \right]$. In the (spatially) flat gauge that we are using ($\psi_2 = E_2 = 0$) this first term is absent. The third term in (\ref{start}) is 
the term $ \zeta_{{\rm exact},2} \vert_{{\rm sourced \; by \; } \delta_1 \varphi \; {\rm and } \;  \delta_1 g_{\mu \nu}} $ that we are disregarding. Therefore
\begin{equation}
\zeta_{{\rm exact},2}  \vert_{{\rm sourced \; by \; } \delta_1 A_\mu}  =  - \frac{\sH}{\rho_0'} \delta_2 \rho  \vert_{{\rm sourced \; by \; } \delta_1 A_\mu}  
\label{zexact-2-formal}
\end{equation}
We stress that this is an exact relation. 

From now on, when we write a second order quantity we only mean the part sourced by $\delta_1 A_\mu$, without indicating it explicitly. The quantity $\delta_2 \rho$ in (\ref{zexact-2-formal}) is the second order perturbation of $-T^0_0$. By evaluating it, we have
\begin{equation}
\zeta_{{\rm exact},2}  =  - \frac{\sH}{\rho_0'} \left[ \frac{1}{a^2} \, \varphi_0' \, \delta_2 \varphi' 
- \frac{1}{a^2} \varphi_0^{' 2} \, \phi_2 + V_{,\varphi} \, \delta_2 \varphi + \frac{1}{2} \left( E^2 + B^2 \right)
\right]
\end{equation}
where $\phi_2$ is the metric perturbations defined in (\ref{deco-metric}), while $E$ and $B$ are  the electric and the magnetic field modes, respectively. Using  eq. (\ref{phi2}) to eliminate $\phi_2$ from this expression, we find
\begin{equation}
\zeta_{{\rm exact},2} = \zeta_2 - \frac{\zeta_2'}{3 \, {\cal H}} - \frac{S^{(3)}}{3 \, {\cal H} } + \frac{E^2+B^2}{6 \left( \rho^{0} + p^{0} \right)}
\label{zexact-2-result}
\end{equation}
where $S^{(3)}$ is the quantity defined in (\ref{S2-S3}), while $\rho_0$ and $p_0$ are  the background energy density and pressure, respectively. Also this relation is exact.

We now show that the last three terms on the right hand side of this expression can be completely disregarded with compared to the first one. We do so by showing that (1) they are already subdominant during inflation, and (2) they decrease relatively to the first term by many orders of magnitude during reheating. Already the statement (1) would be sufficient to disregard them.

To verify the statement (1), we compare the r.m.s. of the various terms during inflation, when the modes are on super horizon scales. From (\ref{zetavar}), we have
\begin{equation}
\sqrt{ \left\langle \zeta_2^2 \right\rangle } \simeq 4 \, {\cal P} \, \ln^2 \left( \frac{a \left( \tau \right)}{a_{\rm in}} \right)
\end{equation}
where we remind that ${\cal P}$, defined in eq. (\ref{power-std}),  is the standard result for the first order power spectrum. We then have
\begin{equation}
\frac{\zeta_2'}{3 {\cal H} \zeta_2} \vert_{\rm r.m.s.} \simeq \frac{2}{3 \, \ln \frac{a \left( \tau \right)}{a_{\rm in}}} \;\;\;\;\;\; {\rm during \; inflation}
\end{equation}
This ratio evaluates to $2/ \left( 3 N_{\rm tot} \right) < 0.01$ at the end of inflation (we recall that $N_{\rm tot}$ denotes the total number of e-folds of inflation).

For the third term in (\ref{zexact-2-result}), we see that the quantity $S^{(3)}$  defined in (\ref{S2-S3})
has three terms that are parametrically of the same order. Therefore we can estimate
\begin{equation}
S^{(3)}  \vert_{\rm r.m.s.} \lsim \frac{3 \, I^2}{2  a^2 M_p^2} \, \left\langle \delta A_i' \, \delta A_i \right\rangle
\end{equation}
where the factor of $3$ accounts for the possibility that the contributions from the three terms add up
in magnitude, although there may actually be cancellations (in this way we obtain a safe upper bound
for  this third term). Inserting (\ref{A-deco}) in this expression, and using (\ref{full-sol}) in the super horizon regime (we recall that $n=2$), we obtain
\begin{equation}
\frac{S^{(3)}}{3 {\cal H} \zeta_2} \vert_{\rm r.m.s.} \lsim  \frac{3 \, \epsilon}{2 \,  \ln \frac{a \left( \tau \right)}{a_{\rm in}}} \;\;\;\;\;\; {\rm during \; inflation}
\end{equation}
So we see that the contribution of the third term  in (\ref{zexact-2-result}) is suppressed by an  $\epsilon$ factor with respect to the already negligible contribution from the second term.


For the last term in (\ref{zexact-2-result}), using eq. (\ref{rho-EB}), we have instead
\begin{equation}
\frac{E^2+B^2}{6 \left( \rho^{0} + p^{0} \right) \zeta_2} \vert_{\rm r.m.s.} 
\simeq \frac{3}{4 \, \ln \frac{a \left( \tau \right)}{a_{\rm in}}} \;\;\;\;\;\; {\rm during \; inflation}
\end{equation}
which again evaluates to $< 0.01$ at the end of inflation.

We see that the corrections to $\zeta_{{\rm exact},2} - \zeta_2$ can be disregarded already at the end of inflation (they are smaller than the accuracy with which we have evaluated $\zeta_2$). Although this is not needed, we can actually verify that these corrections even decrease by several orders of magnitude during reheating. 

For a massive inflaton potential, $\vert \varphi_0 \vert \propto a^{-3/2}$ during reheating. Therefore, the energy density and pressure of the inflaton behave as those of non-relativistic matter. The energy density in the gauge field instead decreases as $a^{-4}$ at the super-horizon scales of our interest.
Therefore, the system rapidly approaches the single fluid regime, with a frozen $\zeta_{{\rm exact},2} \simeq \zeta_2$. One can easily verify that the last two terms in (\ref{zexact-2-result}) also rapidly decrease; specifically, they scale as $a^{-3/2}$ and $a^{-1}$, respectively. Assuming an instantaneous inflaton decay at $t=t_{\rm reh}$, the ratio of the scale factor between the end of inflation and reheating is
\begin{equation}
\frac{a_{\rm end \; infl}}{a_{\rm reh}} \simeq 10^{-10} \, \left( \frac{T_{\rm rh}}{10^9 \, {\rm GeV}} \right)^{4/3} \, \left( \frac{10^{15} \, {\rm GeV}}{H_{\rm inf}} \right)^{2/3}
\end{equation}
where $T_{\rm rh}$ is the temperature of the bath formed by the inflaton decay products.

Therefore, we have explicitly verified that  $\zeta = - \frac{H}{\dot{\varphi}_0} \, \delta \varphi$ is a perfectly good expression for the gauge invariant curvature perturbation in this model.

\section{The In-In Formalism}

In this Appendix we verify that the Green's function method employed in the text is equivalent to the ``in-in'' formalism at leading order.  
To compute correlation functions using the in-in method, we must first identify the interaction Hamiltonian.  To this end, we employ the 
Arnowitt-Deser-Misner (ADM) form of the metric and integrate out the lapse function and shift vector.  This procedure has been described in 
\cite{Seery:2008ms} and also \cite{Barnaby:2011vw}, so we do not reproduce the details of the calculation here.  The quadratic action for the 
inflaton perturbation is
\begin{equation}
\label{S2}
 S_2 = \frac{1}{2}\int d\tau d^3x a^2 \left[ (\partial_\tau\delta\varphi)^2 - \partial_i\delta\varphi\partial_i\delta\varphi - \left(a^2 V'' - 3 \frac{(\varphi_0')^2}{M_p^2}\right)\delta\varphi^2 \right] \, .
\end{equation}
The cubic interaction terms in the Lagrangian are
\begin{eqnarray}
  S_3 &=& \int d\tau d^3x a^4 \frac{I_{,\varphi}}{I} \delta \varphi \left[   \vec{E}^2 - \vec{B}^2 \right]  \label{S3} \\
      &+& \int d\tau d^3x a^4 \frac{\varphi_0'}{\sH M_p^2} \delta\varphi \left[ -\frac{1}{4}(\vec{B}^2 + \vec{E}^2) 
            - \frac{1}{2a^4}\Lap^{-1}\partial_\tau \left( a^4 \vec{\nabla}\cdot (\vec{E}\times \vec{B}) \right)     \right]\nonumber
\end{eqnarray}
Here we have suppressed terms of the form $(\delta\varphi)^3$, which are irrelevant for our calculation.  Varying (\ref{S2}) and (\ref{S3}) 
and expanding $\delta\varphi = \delta_1\varphi + \delta_2\varphi$ reproduces exactly the master equation (\ref{masterapp}).  

For the choice of $I(\varphi)$ under consideration the interactions on the first line of (\ref{S3}) are controlled by the dimensionful 
coupling $\left|\frac{I_{,\varphi}}{I}\right| \sim \sqrt{\frac{1}{\epsilon}}\frac{1}{M_p}$.  In contrast, the interactions in the second 
line of (\ref{S3}) are controlled by a coupling $\left|\frac{\varphi_0'}{\sH M_p^2}\right| \sim \frac{\sqrt{\epsilon}}{M_p}$.  Therefore in 
the slow roll limit, $\epsilon\ll 1$, the interactions on the first line of (\ref{S3}) are the dominant ones.

At leading order in a slow roll expansion, the interaction Hamiltonian $H_I(t) = -\int d^3x a^3 \mathcal{L}_3$ can be written as
\begin{equation}
 H_I(t) = \frac{\dot{\varphi}_0}{H} \int d^3 q J_{\vec q}(\tau) \zeta_{-{\vec q}}(\tau) \, ,
\end{equation}
where $\zeta = -\frac{H}{\dot{\varphi}_0}\delta\varphi$ is the curvature perturbation and the source $J_k(\tau)$ was defined in (\ref{J-F}).  The in-in formula is
\begin{eqnarray}
  \langle \zeta_{\vec k_1} \zeta_{\vec k_2} \cdots \zeta_{\vec k_n}(t) \rangle &=& 
 \sum_{N=0}^{\infty} (-i)^N \int^t dt_1 \int^{t_1} dt_2 \cdots \int^{t_{N-1}} dt_N \label{in-in} \\
  && \langle \left[ \left[ \left[  \zeta_{\vec k_1} \zeta_{\vec k_2} \cdots \zeta_{\vec k_n}(t) , H_{I}(t_1) \right],H_I(t_2)\right] \cdots , H_I(t_N) \right] \rangle \, .
 \nonumber
\end{eqnarray}
A key simplification arises from noting that the mode functions of the produced gauge fluctuations are real-valued, up to an irrelevant constant phase.  This implies that the produced gauge field fluctuations are commuting variables, to a very good approximation.  We have
\begin{equation}
 \left[\partial_\tau A_i , A_j \right] \approx 0 \, ,
\end{equation}
where it is understood that only superhorizon modes are relevant.  Consequently, the source terms $J_q(t)$ in (\ref{in-in}) are mutually commuting and they may be pulled out of the nested commutator.  The remaining commutators are easily evaluated using the formula
\begin{equation}
 \left[\zeta_{\vec k_1}(\tau_1) , \zeta_{\vec k_2}(\tau_2)\right] 
  \cong -i \frac{H^2}{\dot{\varphi}_0^2} \, \frac{G_{k_1}(\tau_1,\tau_2)}{a(\tau_1)a(\tau_2)} \, \delta^{(3)}\left({\vec k_1}+{\vec k_2}\right) \, , \label{commutator}
\end{equation}
where the Green function was defined in (\ref{GreenFunction}).  This formula is valid only for $\tau_1\geq \tau_2$.

Using the commutativity of the source terms and the formula (\ref{commutator}) it is straightforward to evaluate the sourced contribution to the $n$-point correlation functions of $\zeta$.  For the 2-point and 3-point we have
\begin{eqnarray}
&& \left. \langle \zeta_{\vec k_1}\zeta_{\vec k_2}(\tau) \rangle\right|_{\mathrm{sourced}} \approx 
\left(-\frac{H}{a \dot{\varphi}_0}\right)^2 \int_{-\infty}^\tau d\tau_1d\tau_2 \,
           G_{k_1}(\tau,\tau_1)G_{k_2}(\tau,\tau_2) \,\langle J_{\vec k_1}(\tau_1) J_{\vec k_2}(\tau_2)\rangle \, , \nonumber \\
&& \left. \langle \zeta_{\vec k_1}\zeta_{\vec k_2}\zeta_{\vec k_3}(\tau) \rangle\right|_{\mathrm{sourced}} \approx 
\left(-\frac{H}{a\dot{\varphi}_0}\right)^3 \int_{-\infty}^\tau d\tau_1d\tau_2 d\tau_3
         \,  G_{k_1}(\tau,\tau_1)G_{k_2}(\tau,\tau_2) G_{k_3}(\tau,\tau_3)\nonumber \\
&& \hspace{80mm} \times \langle J_{\vec k_1}(\tau_1) J_{\vec k_2}(\tau_2)J_{\vec k_3}(\tau_3)\rangle \, . \nonumber
\end{eqnarray}
These coincide exactly with what we obtained previously using the Green function method.  We have also verified that the cross-correlation of gauge field fluctuations with the curvature perturbation agrees with what was presented in \cite{Caldwell:2011ra}, at leading order.

\end{document}